%% file: ComplexGeodesics_JHEP.tex
\documentclass[onecolumn,12pt]{article}
\usepackage{jheppub}
\usepackage{url}
\usepackage[font={small}]{caption}
\usepackage{subfigure}
\usepackage{graphicx}
\usepackage{dcolumn}
\usepackage{float}
\usepackage[normalem]{ulem}
\usepackage{mathrsfs}
\usepackage{multicol}
\usepackage{cancel}
\usepackage{mathtools}
\usepackage{hyperref}

\usepackage{subfigure}
\usepackage{xcolor}
\def\({\left(} \def\){\right)}
\def\[{\left[} \def\]{\right]}

\newcommand{\ie}{{\it i.e.,}\ }

\newcommand{\bea}{\begin{eqnarray}}
\newcommand{\eea}{\end{eqnarray}}

\renewcommand{\eqref}[1]{(\ref{#1})}

\definecolor{crimson}{rgb}{0.86, 0.08, 0.24}
\definecolor{darkpastelgreen}{rgb}{0.01, 0.75, 0.24}

\begin{document}


\begin{flushright}
QMUL-PH-22-37\\
\end{flushright}

\title{Complex geodesics in de Sitter space}

\author{Shira Chapman$^1$, Dami\'an A. Galante$^2$, Eleanor Harris$^2$, Sameer U. Sheorey$^2$ and David Vegh$^3$}

\affiliation{$^1$ Department of Physics, Ben-Gurion University of the Negev, Beer Sheva 84105, Israel}
\affiliation{$^2$ Department of Mathematics, King’s College London, the Strand, London WC2R 2LS, UK}
\affiliation{$^3$ Centre for Theoretical Physics,
Queen Mary University of London,  London E1 4NS, UK}

\emailAdd{schapman@bgu.ac.il}
\emailAdd{damian.galante@kcl.ac.uk}
\emailAdd{eleanor.k.harris@kcl.ac.uk}
\emailAdd{sameer.sheorey@kcl.ac.uk}
\emailAdd{d.vegh@qmul.ac.uk}

\abstract{
\sloppy
The two-point function of a free massive scalar field on a fixed background can be evaluated in the large mass limit by using a semiclassical geodesic approximation. In de Sitter space, however, this poses a puzzle. Certain spacelike separated points are not connected by real geodesics despite the corresponding two-point function in the Bunch-Davies state being non-vanishing. We resolve this puzzle by considering complex geodesics after analytically continuing to the sphere. We compute one-loop corrections to the correlator and discuss the implications of our results to de Sitter holography.}

\setcounter{tocdepth}{2}

\maketitle

\section{Introduction}
\input{sections/intro.tex}

\section{The scalar two-point function in dS}
\label{sec_twopoint}
\input{sections/twopt.tex}

\clearpage

\section{The geodesic approximation} \label{sec:geo}
\input{sections/geo.tex}

\section{Euclidean two-point functions} \label{sec:sphere}
\input{sections/geosphere.tex}
\section{Lorentzian two-point functions} \label{lorentzian_props}
\input{sections/lorentz2pt.tex}

\section{Outlook} \label{sec:outlook}
\input{sections/outlook.tex}

\section*{Acknowledgements}

We would like to acknowledge Dionysios Anninos, Alice Bernamonti, Jackson Fliss, Ben Freivogel, Federico Galli, Diego Hofman,  Mir Mehedi Faruk, Rob Myers, K\'evin Nguyen, Jake Phillips, Shan-Ming Ruan, Jan Pieter van der Schaar and Manus Visser for useful discussions.

The work of S.C. is supported by the Israel Science Foundation (grant No. 1417/21), the German Research Foundation through a German-Israeli Project Cooperation (DIP) grant ``Holography and the Swampland'' and the Alon Fellowship for young faculty members.  S.C. further acknowledges the support of Carole and Marcus Weinstein through the BGU Presidential Faculty Recruitment Fund.
The work of D.A.G. is funded by UKRI Stephen Hawking Fellowship ``Quantum Emergence of an Expanding Universe". E.H. is funded by the STFC studentship “Aspects of black hole and cosmological horizons”. S.U.S. is funded by the Royal Society under the grant ``The Resonances of a deSitter Universe". D.V. is funded by the STFC Ernest Rutherford grant ST/P004334/1.

D.A.G. would like to further thank the University of Amsterdam, the University of Kentucky and the Perimeter Institute for kind hospitality during the completion of this work. Research at Perimeter Institute is supported by the Government of Canada through the Department of Innovation, Science and Economic Development and by the Province of Ontario through the Ministry of Colleges and Universities.

  \appendix

\section{WKB approximation}
\label{app_WKB}
\input{sections/appwkb.tex}

\section{Quantum mechanical path integral} \label{path_integral_app}
\input{sections/apppathintegral.tex}

\section{Geodesic equation in higher dimensions}
\label{app_ddim}
\input{sections/appd-dimensionaleom.tex}

\section{Details on the stretched horizon} \label{app:details}
\input{sections/appdetails.tex}

\bibliographystyle{JHEP}

\bibliography{bibliography}

\end{document}

%% file: sections/intro.tex
Geodesics play an important role in our understanding of curved spacetimes, both at a classical and semiclassical level. For instance, it is known that correlation functions of quantum field theories in fixed curved backgrounds are related to geodesics. In the worldline formalism \cite{Schubert:2001he, Bastianelli:2005rc}, one can schematically compute $G(X, Y)$, the correlator of a free massive scalar field between two points $X$ and $Y$, as a path integral,
\begin{equation}\label{geoworld}
G(X,Y) = \int D{\mathcal{P}} \, e^{-m L[\mathcal{P}]}  \approx \sum_{g \, \in\, {\text{geodesics}}} e^{-m L_g}\,,
\end{equation}
where $m$ is the mass of the scalar field, $\mathcal{P}$ is a path connecting $X$ and $Y$, and $L[\mathcal{P}]$ is the length of that path. In the large mass limit, it is possible to take a saddle point approximation that reduces the path integral to a sum over geodesic lengths $L_g$. These discussions go back to the seminal work by Bekenstein and Parker \cite{PhysRevD.23.2850}. See also \cite{PhysRevD.19.438, parker_toms_2009}.

This formula might look surprising since it is well known \cite{hawking_ellis_1973, jacobson} that not all points in a Lorentzian spacetime can be connected by geodesics.\footnote{By the upper semi-continuity of arc-length, spacelike geodesics in Lorentzian manifolds always have (locally) maximal lengths \cite{Wald:1984rg}. However,  it is possible to find  scenarios where curve lengths connecting two points are unbounded from above, and in those cases, geodesics do not exist at all.}
For instance, in de Sitter (dS) space certain spacelike separated points are not connected by a geodesic. The global metric of dS$_d$ is given by
\begin{equation}
\frac{ds^2}{\ell^2} = -dT^2 + \cosh^2 T d\Omega_{d-1}^2 \,, \label{global_dS}
\end{equation}
where $T \in \mathbb{R}$ is the global time and $d\Omega_{d-1}^2$ is the metric on the unit $(d-1)$-sphere. The de Sitter length scale $\ell$ will be set to one from now on. An inertial observer in this spacetime does not have access to the full geometry and is instead confined inside a cosmological event horizon. The causal region that such an observer has access to is described by the static patch metric,
\begin{equation} \label{static_metric}
ds^2 = - (1-r^2) dt^2 + \frac{dr^2}{1-r^2} + r^2 d\Omega_{d-2} \,,
\end{equation}
where $t \in \mathbb{R}$ and $ 0\leq r \leq 1$ for $d>2$, while $-1 \leq r \leq 1$ for $d=2$.  Early work on static patch correlation functions includes \cite{Balasubramanian:2002zh, Goheer:2002vf, Anninos:2011af}. Geodesics in dS$_2$ have been recently studied in \cite{Chapman:2021eyy, Galante:2022nhj}. Extremal surfaces in dS (including geodesics in $d=3$) have been studied in \cite{Fischetti:2014uxa}, while other extended objects such as Wilson lines have been studied in \cite{Castro:2020smu}. The geodesic approximation for late time correlators in dS$_3$ has recently been studied in \cite{Hikida:2022ltr}.

As mentioned above, in dS there are no real geodesics connecting certain spacelike separated points. However, at those points the two-point correlator for a free massive scalar field in the dS invariant, Bunch-Davies (or Euclidean) state is well-defined. In fact, this correlator is known analytically for all points and all spacetime dimensions \cite{Spradlin:2001pw, Anninos:2012qw}. The two-point function between points $X$ and $Y$ in dS only depends on the dS invariant distance between the two points, which we call $P_{X,Y}$. In the large mass limit, we find that
\begin{equation}
G(P_{X,Y}) \approx \frac{m^{\frac{d-3}{2}}}{2 (2\pi) ^{\frac{d-1}{2}}} \left[ \frac{   e^{-m \cos ^{-1}P_{X,Y}}}{\left(1-P_{X,Y}^2\right)^{\frac{d-1}{4}}}  +  \frac{ \left(-1\right)^{\frac{d-1}{2}} e^{-m (2\pi - \cos ^{-1}P_{X,Y})}}{(1-P_{X,Y}^2)^{\frac{d-1}{4}}} \right] \,, \label{gintro}
\end{equation}
where this asymptotic form is valid for all points satisfying 
  \begin{equation}\label{p condition}
   |1-P_{X,Y}^2| \gtrsim m^{-2}\;.
   \end{equation}

If $P_{X,Y} >-1$, there is always a geodesic connecting $X$ and $Y$, and its length is given by $L_g = \cos^{-1} P_{X,Y}$.\footnote{Points with $-1<P_{X,Y} <1$ are spacelike separated, while for $P_{X,Y}>1$ points become timelike separated.  In our conventions, timelike separated points have imaginary geodesic length.
} In this case, the second term in \eqref{gintro} is always exponentially suppressed in the large mass limit and can be neglected, and so the correlator takes the form in equation \eqref{geoworld}, as expected. The term in the denominator can be computed from quadratic fluctuations around the geodesic length.

If $P_{X,Y} <-1$, on the other hand, the two points are spacelike separated, but there are no real geodesics connecting them. In that case, both terms in (\ref{gintro}) contribute to the correlator in the large mass limit, and it naively seems as if the geodesic length becomes complex. But spacelike geodesics have real length, so in this case it is not clear how to reconcile the prescription in equation \eqref{geoworld} with the fact that geodesics do not exist between such points.

In this paper, we solve this apparent tension by considering Euclidean geodesics. The analytic continuation of dS to Euclidean signature is given by the sphere. There always exist two geodesics that form a great circle between any two points on the sphere. We compute the lengths of these geodesics as well as the one-loop correction to the Euclidean correlator coming from quadratic fluctuations of the geodesic length.

We show that for $P_{X,Y} <-1$, it is necessary to keep the contributions from both (Euclidean) geodesics even though on the sphere only one has the shortest length. Upon analytically continuing back to Lorentzian signature, we reproduce the precise form of (\ref{gintro}), up to an overall coefficient. There exist other geodesics that wrap around the great circle more than once, but these are always suppressed in the large mass limit. Our results are valid in any number of spacetime dimensions and for any choice of points in dS provided that \eqref{p condition} holds.

The remainder of the text is organised as follows. In section \ref{sec_twopoint}, we review the computation of the Wightman correlator in dS and compute its large mass expansion. In section \ref{sec:geo}, we review the calculation of Lorentzian geodesics in dS, recovering the result that not all spacelike separated points in dS are connected by a real geodesic. In section \ref{sec:sphere}, we move to the sphere and compute the Euclidean correlator in the geodesic approximation. We also compute one-loop corrections around each saddle point. In section \ref{lorentzian_props}, we analytically continue this result to Lorentzian signature to obtain the correct Lorentzian correlator. We apply our findings to several examples of timelike and spacelike geodesics, including those between opposite  stretched horizons, \ie surfaces of fixed $r$ in (\ref{static_metric}). This leads into section \ref{sec:outlook}, where we discuss these results in the light of recent developments regarding dS space and holography. We relegate some technical details to appendices. In Appendix \ref{app_WKB}, we recover the asymptotic form of the two-point correlator from a WKB approximation. In Appendix \ref{path_integral_app}, we solve the quantum mechanical path integral needed to compute one-loop corrections to the Euclidean correlator. In Appendix \ref{app_ddim}, we discuss sphere geodesics in $d\geq 2$. Finally,  in Appendix \ref{app:details} we give details on the computation of correlations between stretched horizons.

\

\noindent \textbf{Note added:}
During the preparation of this paper we became aware of  \cite{lars}, which presents a related discussion on the role of complex geodesics in de Sitter space.

%% file: sections/twopt.tex
Consider the action for a free massive scalar field in dS$_d$,
\begin{equation}
S_\phi = -\frac{1}{2} \int d^dx \sqrt{-g} \left[ g^{\mu\nu} \partial_\mu \phi \partial_\nu \phi  +m^2 \phi^2 \right] \,,
\end{equation}
where $m$ is the mass of the scalar field and $g_{\mu \nu}$ is the $d$-dimensional de Sitter metric. We will study the Wightman two-point function in the Bunch-Davies (or Euclidean) vacuum state $| E \rangle$ \cite{Spradlin:2001pw, Anninos:2012qw}, $G(X,Y) = \langle E | \phi(X) \phi(Y) |  E \rangle$, where $X$ and $Y$ are arbitrary points on global de Sitter. This two-point function is a solution to the Klein-Gordon equation in dS$_d$,
\begin{equation}
(\square - m^2) G(X,Y) = 0 \,.
\end{equation}
Given that the Euclidean state is invariant under the dS isometries, the two-point function can only depend on the two points through their de Sitter invariant length $P_{X,Y}$. This quantity can be easily defined in embedding space as
\begin{equation}
	P_{X,Y} \equiv \eta_{IJ} X^I Y^J , \quad \eta_{IJ} = \text{diag} \overbrace{(-1, 1, \ldots, 1)}^{d + 1}  \,. \label{PXY}
\end{equation} 
Note that in this last expression $X$ and $Y$ are coordinates describing the embedding of the $d$-dimensional de Sitter hyperboloid inside $(d+1)$-dimensional Minkowski space. For instance, if we choose to parameterise the hyperboloid with global coordinates, the de Sitter invariant length is given by
\begin{equation} \label{P_global12}
	P_{X,Y} = - \sinh T_X \sinh T_Y + \cosh T_X \cosh T_Y \sum_{i = 1}^d \omega^i_X {\omega}^i_Y \,,
\end{equation}
where $\omega^i$ are coordinates on the $(d-1)$-sphere, \ie $\sum_{i=1}^{d} \left(\omega^i \right)^2 = 1$. 

It is interesting to note that \cite{Spradlin:2001pw}
\begin{equation}
\begin{cases}
P_{X,Y}  >1 \, ,  &\text{for timelike separated points} \,; \\
P_{X,Y}  =1  \, ,  &\text{for coincident or null separated points} \,; \\
P_{X,Y}  <1  \, ,  &\text{for spacelike points} \,; \\
P_{X,Y}  =-1 \, ,  &\text{when $X$ is null separated from the antipodal point of $Y$} \,; \\
P_{X,Y}  <-1 \,  ,  &\text{when $X$ is timelike separated from the antipodal point of $Y$} \,.
\end{cases} \nonumber
\end{equation}
Going back to the two-point function, if we write $G(X,Y) = G(P_{X,Y})$, then the Klein-Gordon equation becomes,
\begin{equation}
(1-P_{X,Y}^2) \partial_{P_{X,Y}}^2 G(P_{X,Y}) - d P_{X,Y} \partial_{P_{X,Y}} G(P_{X,Y}) - m^2 G(P_{X,Y}) = 0 \,. \label{kg_P}
\end{equation}

The unique solution to this hypergeometric equation that correctly reproduces the expected short distance behaviour of the two-point function (and does not have singularities at antipodal points) is given by
\begin{equation}
G(P_{X,Y}) = \frac{\Gamma(h_+) \Gamma(h_-)}{(4\pi)^{d/2} \Gamma \left( \frac{d}{2}\right)} \, _2F_1 \left( h_+, h_-; \frac{d}{2}; \frac{1+P_{X,Y}}{2} \right)\, , \,\,  h_\pm = \frac{(d-1)}{2} \pm \sqrt{\left(\frac{d-1}{2}\right)^2-m^2} \,. \label{exact_2pt}
\end{equation}
Note that, by definition, this is also the two-point function obtained from analytically continuing the two-point function on $S^d$ to Lorentzian dS spacetime.

We are interested in the large mass expansion of the above correlator. The asymptotic form of the hypergeometric function when some of the parameters are large is non-trivial. The expansion of interest in our case is the one where the first two parameters become large in the imaginary direction (and with opposite signs). Asymptotic expressions in this limit were first found in \cite{watson}. See also \cite{Cvitkov}  for the latest state of affairs.

There are two different large mass expansions for the hypergeometric function, depending on whether its last argument is positive or negative. When $P_{X,Y} <-1$, the form of the correlator in the large mass limit is given by  \cite{watson}
\begin{equation}
\begin{split}
G(P_{X,Y}) & \approx   \frac{m^{\frac{d-3}{2}}}{2(2\pi) ^{\frac{d-1}{2}}} \left[ \frac{   e^{-m \cos ^{-1}P_{X,Y}}}{\left(1-P_{X,Y}^2\right)^{\frac{d-1}{4}}}  +  \frac{ \left(-1\right)^{\frac{d-1}{2}} e^{-m (2\pi - \cos ^{-1}P_{X,Y})}}{(1-P_{X,Y}^2)^{\frac{d-1}{4}}} \right] \,,\label{Pminus1} \\
& =  \frac{ m^{\frac{d-3}{2}}}{(2\pi) ^{\frac{d-1}{2}} } \, \textrm{Re}\left[ {e^{-m \cos^{-1} P_{X,Y} }\over (1-P_{X,Y}^2)^{\frac{d-1}{4}}} \right] \quad , \quad P_{X,Y} < -1 \,. 
\end{split}
 \end{equation}
In the first line, the correlator is written in a form inspired by the geodesic approximation. In the second line the correlator is manifestly real, consistent with the fact that the points are spacelike separated. 

When $P_{X,Y} > -1$, the second term in (\ref{Pminus1}) is always exponentially suppressed in the large mass limit, so the correlator takes the form
\begin{equation}
\begin{split}
G(P_{X,Y}) \approx \frac{m^{\frac{d-3}{2}}}{2(2\pi) ^{\frac{d-1}{2}}} \frac{ \exp  \left( -m \cos ^{-1}P_{X,Y}\right)}{\left(1-P_{X,Y}^2\right)^{\frac{d-1}{4}}}  \, , \,\,\, P_{X,Y} >-1 \,. 
\end{split}
\label{P_greater}
 \end{equation}
For $-1<P_{X,Y}<1$, it is straightforward to check that the correlator is real. When points become timelike separated, then $P_{X,Y}>1$ and the expression becomes manifestly complex, consistent with the fact that, in our conventions, timelike geodesics have an imaginary length. 

Both (\ref{Pminus1}) and (\ref{P_greater}), can be obtained by solving the Klein-Gordon equation (\ref{kg_P}) in a WKB expansion, which provides an independent check of these asymptotic expansions; see Appendix \ref{app_WKB}. 

It is clear that both approximations break down when $P_{X,Y} \sim \pm 1$. When $P_{X,Y}$ is close to one, the correlator is chosen to mimic the short distance singularity in flat space \cite{Spradlin:2001pw}, so it takes the form 
  \begin{equation}
G (P_{X,Y} \sim 1)  \approx   \frac{\Gamma \left(\frac{d}{2}\right)}{(2 \pi )^{\frac{d}{2}} (d-2)} \frac{1}{(1-P_{X,Y})^{\frac{d}{2}-1}}  \,, \label{P1}
\end{equation}
 which is independent of $m$.\footnote{Note that in $d=2$ the correlator actually diverges logarithmically as $G (P_{X,Y} \sim 1)  \approx -\log \left(1-P_{X,Y}\right)/4 \pi$, which is consistent with the expected QFT behaviour.}
When $P_{X,Y}=-1$, the last argument in the hypergeometric function is zero, so in the large mass limit we obtain
 \begin{equation}
G (P_{X,Y}=-1)  \approx   \frac{m^{d-2}}{\Gamma(d/2) 2^{d-1} \pi^{d/2-1}} e^{-m \pi} \,. \label{t_zero_d}
\end{equation}
It is easy to see that these limits do not commute with taking the large mass limit first.

%% file: sections/geo.tex
So far, we have obtained the two-point correlator in the large mass limit by finding the asymptotic form of the relevant hypergeometric function. A WKB approach yields the same answer, see Appendix \ref{app_WKB}. The main purpose of this work is to reconcile these results with the expression coming from the geodesic approximation,
\begin{equation} \label{geo_approx_eq}
G(P_{X,Y}) = \int D{\mathcal{P}} e^{-m L[{\mathcal{P}}]}  \approx \sum_{g \, \in \, {\text{geodesics}}} e^{-m L_g}\,,
\end{equation}
even when real geodesics do not exist. Here, the path integral is over all possible paths $\mathcal{P}$ connecting the points $X$ and $Y$, and $L[{\mathcal{P}}]$ is the length of that path. In the large mass limit, the path integral can be approximated in a saddle point approximation by computing the geodesic length $L_g$ connecting the two points. If there is more than one geodesic, we need to sum over them appropriately.

We start by reviewing how to compute geodesics in dS in cases where real geodesics do exist. For simplicity, we demonstrate this in $d=2$, but the results can be generalised to higher dimensions.

\subsection{Review of real geodesics in dS$_2$} \label{sec_geo}
The global metric of dS$_2$ is given by,
\begin{equation}
ds^2 = -dT^2 + \cosh^2 T \, d\varphi^2 \,,
\end{equation}
with $\varphi \in \left[-\tfrac{\pi}{2}, \tfrac{3\pi}{2} \right)$ and $T \in \mathbb{R}$. 
The length functional is given by
\begin{equation} \label{LorentzianLength}
	L = \int ds = \int d\lambda \, \mathcal{L} (T, \dot{T}, \varphi, \dot{\varphi}, \lambda) = \int d \lambda \sqrt{ (- \dot{T}^2 + \cosh^2 T \dot{\varphi}^2)} \,  ,
\end{equation}
where the dots represent derivatives with respect to the parameter $\lambda$ along the geodesic. Recall that we are working in a slightly unusual convention where timelike geodesics will have a complex length.
The Lagrangian $\mathcal{L}$ does not depend explicitly on $\varphi$, so we can define the following conserved quantity
\begin{equation} \label{phieqn}
	\frac{\partial \mathcal{L}}{\partial \dot{\varphi}} \equiv Q =  \dot{\varphi}  \cosh^2 T  \left( -\dot{T}^2 + \cosh^2 T \dot{\varphi}^2\right)^{-1/2} \,.
\end{equation} 
Since the length functional is invariant under reparametrisation, we may select $\lambda$ such that it is an affine parameter, 
\begin{equation}  \label{lagrangian}
	\mathcal{L}^2=\left( \frac{ds}{d\lambda} \right)^2 = \pm 1 = - \dot{T}^2 + \cosh^2 T \dot{\varphi}^2 \, ,
\end{equation}
where the $\pm$ depends on whether the geodesic is spacelike or timelike, respectively. 
The first order equations (\ref{phieqn}) and (\ref{lagrangian}) can be integrated to find the trajectories of the geodesics, which read 
\begin{eqnarray}
		\tan ( \varphi + \tilde{\varphi} ) &=&  \frac{Q \sinh T }{\sqrt{Q^2 -  \cosh^2 T}} \, , \label{lorenztiansol} 
\end{eqnarray}
where $\tilde{\varphi}$ is a constant of integration which, as well as $Q$, can be determined by the choice of the endpoints of the geodesic.

The length of the geodesic can be expressed in terms of its endpoints at $(T_1, \varphi_1)$ and $(T_2, \varphi_2)$ as
\begin{equation}\label{lenTphi}
	L_g = \cos^{-1} P_{X,Y} =  \cos^{-1} \left[\cosh T_1 \cosh T_2 \cos (\varphi_2 - \varphi_1) -  \sinh T_1 \sinh T_2 \right] \, .
\end{equation}
Timelike separated points have $P_{X,Y}>1$, and their geodesic length is complex. For example, geodesics between points  $(T_1, \varphi_0)$ and $(T_2, \varphi_0)$ have vanishing conserved charge, $Q=0$, and so their length is given by $L_g = i |T_2-T_1|$.

\begin{figure}[H]
        \centering
         \subfigure[$T_1=0$]{
                \includegraphics[height=5cm]{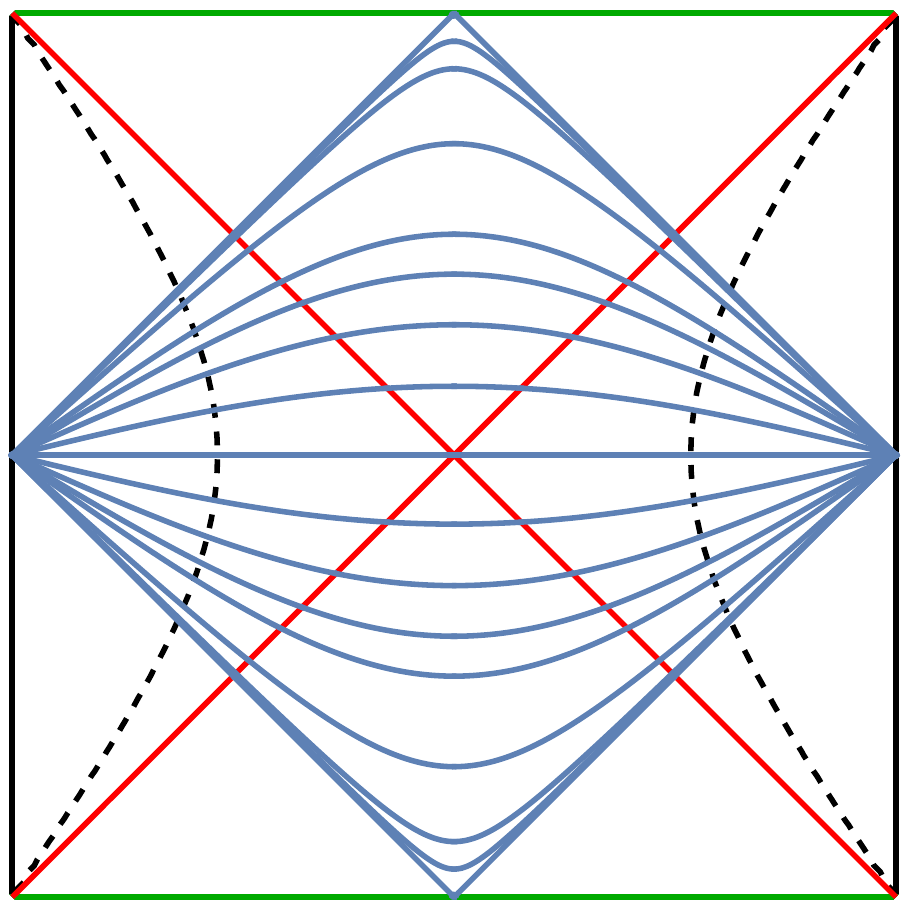}\label{fig:ds0}}  \qquad \qquad
        \subfigure[$T_1=1$]{
                \includegraphics[height=5cm]{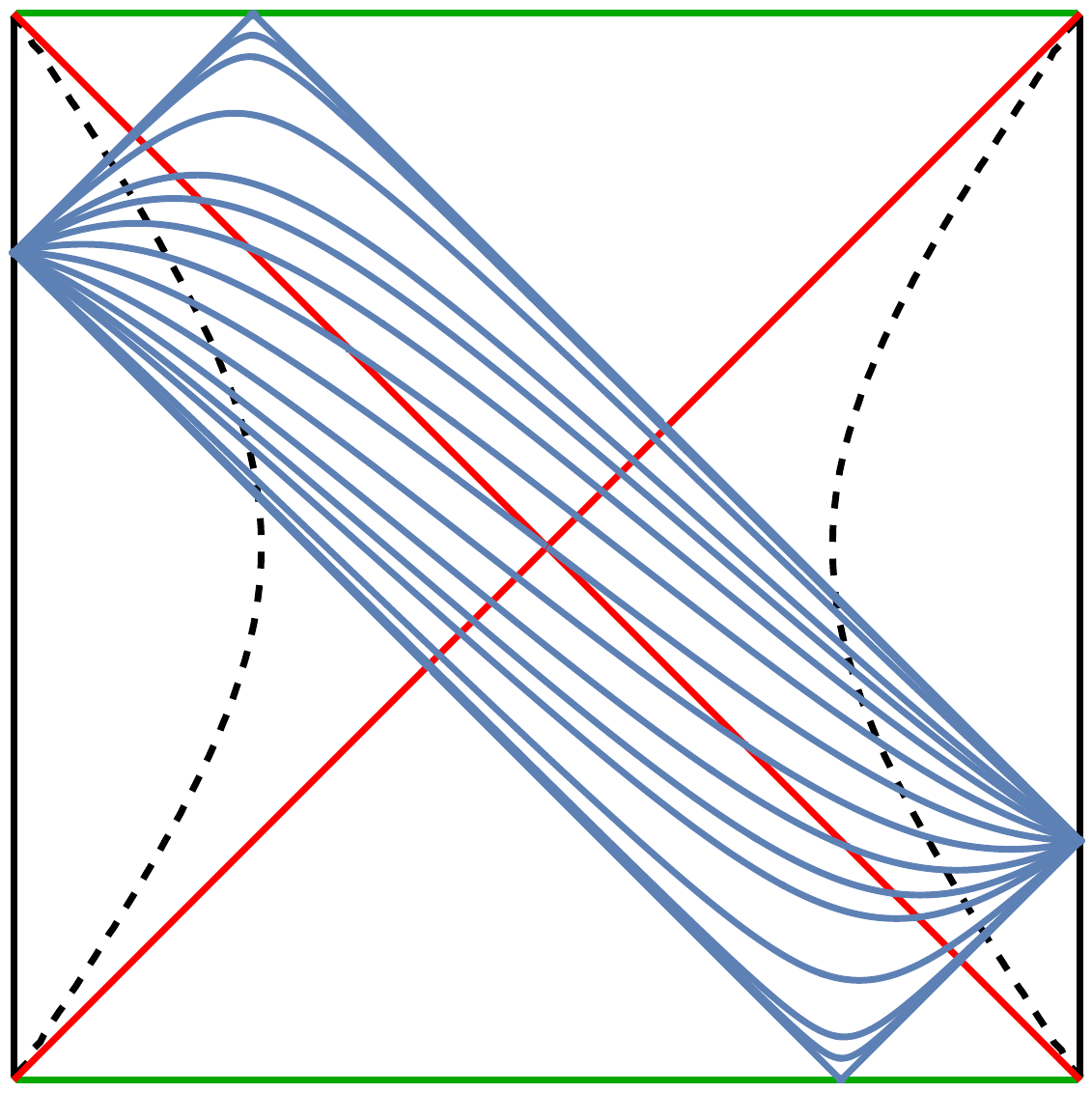} \label{fig:ds1}}  \qquad
                 \caption{Penrose diagram for (half of) dS$_2$ with geodesics in blue. The horizons are drawn in red, and past and future infinity are in green. In dashed black, we also plot the position of stretched horizons.}\label{fig:penroseperlim}
\end{figure}

On the other hand, spacelike separated points have $P_{X,Y}<1$, but real geodesics only exist for $-1\leq P_{X,Y}<1$. 
As an example, consider points on opposite sides of the spatial circle, \ie  $(T_1, \varphi_0)$ and $(T_2, \varphi_0+\pi)$, for which $P_{X,Y}\leq -1$. Requiring equation \eqref{lenTphi} to be real imposes that $T_1=-T_2$ and $P_{X,Y} = -1$. In this case, we have a one-parameter family of geodesics whose charges are given by $|Q|>\cosh{T_1}$. All of them have geodesic length $L_g = \pi$ \cite{Chapman:2021eyy, Jorstad:2022mls}. See figures \ref{fig:ds0} and \ref{fig:ds1}.  This result contrasts with the AdS$_2$ black hole case, where geodesics exist for arbitrarily long times and their length grows linearly with time \cite{Brown:2018bms}.

\subsection{From geodesics to two-point correlators}  \label{sec:tension}
One would like to use the results obtained for the real geodesics in dS to reproduce the form of the correlators studied in section \ref{sec_twopoint}. However, it is clear from the result just shown for points on opposite sides of the spatial circle that real geodesics are not enough.

One can always formally define $L_g = \cos^{-1} P_{X,Y}$, so that (\ref{Pminus1}) becomes
\begin{equation}
G(L_g) \approx \frac{m^{\frac{d-3}{2}}}{2 (2\pi) ^{\frac{d-1}{2}}} \left[ \frac{   e^{-m \, L_g}}{\left(\sin L_g \right)^{\frac{d-1}{2}}}  +  \frac{  e^{-m (2\pi - L_g)}}{\left(\sin (2\pi - L_g) \right)^{\frac{d-1}{2}}} \right] \,. \label{GofL}
 \end{equation} 
For any $P_{X,Y}>-1$, the second term is always exponentially suppressed, and so the propagator can be written in the form of the geodesic approximation  (\ref{geo_approx_eq}). Moreover, we showed that, in those cases, the geodesic length is actually $L_g = \cos^{-1} P_{X,Y}$, both for spacelike and timelike separated points. The only term that needs to be explained is the denominator, that will come from perturbations around the geodesic length.

If $P_{X,Y}<-1$, there seems to be more tension. In this case, we derived that there are no real geodesics connecting these points in dS. The naive continuation of $L_g$ gives a complex geodesic length. Moreover, from the form of (\ref{GofL}), it seems that there are two geodesics contributing to the correlator. As in the previous case, the denominator in each term needs to be explained.

In the next section, we will show that the tension for $P_{X,Y}<-1$ can be cured by looking at geodesics on the sphere.  In all cases, denominators will appear as one-loop corrections to the geodesic length.

Before moving on, let us comment on the special degenerate case of $P_{X,Y} = -1$. In this case, we showed that $L_g = \pi$, which is consistent with $P_{X,Y}=-1$. But we found that there are infinitely many geodesics, which would naively yield an infinite correlator, unless properly regulated. Note also that equation (\ref{P_greater}) diverges in this limit, but that the correct large mass correlator is (\ref{t_zero_d}). In what follows, we will restrict to $P_{X,Y} \neq -1$.

%% file: sections/geosphere.tex
In this section, we compute the two-point correlator on the sphere, where we know that any two points are connected by a geodesic with real (Euclidean) length. We start by focusing on $d=2$.

Considering the sphere is natural since it is the Euclidean continuation of Lorentzian dS spacetime. In fact, if we analytically continue global dS$_2$ using $T \to - i \theta$, the metric in equation (\ref{global_dS}) becomes
\begin{equation}
ds^2 = d\theta^2 + \cos^2 \theta \, d\varphi^2 \,, \label{line_element}
\end{equation}
which is the round metric on $S^2$. Here $\varphi \in [-\pi/2,3\pi/2)$ and $\theta \in [-\pi/2,\pi/2]$. Note that the analytic continuation of the static patch metric also gives the sphere, in a slightly different coordinate system.

It is known that, given two points on the sphere, there is always a great circle that passes through them. The great circle is defined by the intersection of the plane containing the two points and the origin with the sphere. The two segments of the great circle are geodesics connecting the two points.

It will be convenient to use a new set of coordinates $\{ \Theta, \Phi \}$, where the $\Phi$ angle moves around the great circle between the two points, $\Phi \in [0, 2\pi )$ and $\Theta \in [-\pi/2, \pi/2]$. The great circle lies at $\Theta = 0$. Note that you can always move to this coordinate frame for any two points $X$ and $Y$ on the sphere. The metric on this coordinate system is given by
\begin{equation} \label{new_metric}
ds^2 = d\Theta^2 + \cos^2 \Theta \, d\Phi^2 \,,
\end{equation}
so the length functional becomes
\begin{equation} \label{length_funct_2}
\tilde{L} [\Phi_X, \Phi_Y, \Theta] = \int^{\Phi_Y}_{\Phi_X} d\Phi  \sqrt{\dot{\Theta}^2 (\Phi) + \cos^2 \Theta (\Phi)}  \,.
\end{equation}
In this section, we use tildes to denote Euclidean quantities. We would like to compute the Euclidean two-point function on the sphere, using the geodesic approximation, \ie
\begin{equation} \label{euc_path_int}
\tilde{\mathcal{G}}(\Phi_X, \Phi_Y) = \int_{\Theta(\Phi_X)=0}^{\Theta(\Phi_Y)=0} D\Theta(\Phi) \exp (-m \tilde{L}(\Phi_X, \Phi_Y, \Theta)) \,.
\end{equation}
We start by considering geodesics in $d=2$ in section \ref{sec:euclidean_saddles}. We then consider perturbations to the geodesic length in section \ref{sec:pert} and finally, we generalise our results to higher dimensions in section \ref{sec_higher_d}.

\subsection{Euclidean saddle points} \label{sec:euclidean_saddles}

In the large mass limit, (\ref{euc_path_int}) is dominated by its saddle points which are the geodesics connecting $X$ and $Y$. To find these, we extremise the length (\ref{length_funct_2}). The equation of motion stemming from this length functional is
\begin{equation}
	\ddot{\Theta} + 2 \dot{\Theta}^2  \tan \Theta +\sin \Theta \cos \Theta = 0 \,, 
\end{equation}
which is solved by
\begin{equation} \label{theta of phi}
	\Theta ( \Phi) = \pm \sin^{-1} \left[\frac{\sqrt{c_1-1 } \tan (\Phi + c_2)}{\sqrt{1+c_1 \tan^2(\Phi +c_2)}}\right]\,,
\end{equation}
where $c_1$ and $c_2$ are constants of integration. Boundary conditions at the endpoints of integration set $\Theta(\Phi_X) = \Theta(\Phi_Y) = 0$. A generic solution obeying the boundary conditions has $c_1=1$, reducing the solution to $\Theta_{\text{geodesic}} = 0$, \ie the geodesic goes through the great circle, as expected.  Evaluating the action on-shell gives the length of the shorter geodesic,
\begin{equation} \label{L-}
\tilde{L}_{g} = \int^{\Phi_Y}_{\Phi_X} d\Phi  \sqrt{\dot{\Theta}_{\text{geodesic}}^2 + \cos^2 \Theta_{\text{geodesic}}} = \Phi_Y - \Phi_X \,,
\end{equation}
assuming without loss of generality that $0 < \Phi_Y - \Phi_X \leq \pi$. There is also another geodesic that goes around the other side of the great circle and has length
\begin{equation} \label{L+}
\tilde{L}_+ = \int^{2\pi + \Phi_X}_{\Phi_Y} d\Phi  \sqrt{\dot{\Theta}_{\text{geodesic}}^2 + \cos^2 \Theta_{\text{geodesic}}} = 2\pi - (\Phi_Y - \Phi_X) = 2\pi - \tilde{L}_g \,.
\end{equation}

Note that, generically, $\tilde{L}_g \leq \tilde{L}_+$, so only $\tilde{L}_g$ will contribute to the Euclidean correlator in the large mass limit. However, as we will see in section \ref{lorentzian_props}, we require both geodesics in certain cases to reproduce the correct Lorentzian correlator from analytic continuation of the Euclidean result. 

There also exist another (infinite) set of geodesics that wrap multiple times around the great circle. Their lengths are given by  $\pm\tilde{L}_g+ 2\pi n$ with $n \in \mathbb{N}$, and their contribution to the correlator will always be exponentially suppressed in the large mass expansion, even after analytic continuation to Lorentzian spacetime.\footnote{With the exception of $-\tilde{L}_g + 2\pi$, which is actually $\tilde{L}_+$.}

On the sphere, the only case where both $\tilde{L}_g$ and $\tilde{L}_+$ will contribute corresponds to having $\Phi_Y - \Phi_X = \pi$. In this case, there is another set of solutions to (\ref{theta of phi}) that satisfy  the boundary conditions. These are a one-parameter family labeled by $c_1 \in \mathbb{R}^{\geq 1}$ and obtained by setting $c_2 = -\Phi_X$ in (\ref{theta of phi}). These geodesics are given by,
\begin{equation}
\Theta (\Phi) = \pm \sin^{-1} \left[\frac{\sqrt{c_1-1} \tan (\Phi - \Phi_X)}{\sqrt{1+c_1 \tan^2(\Phi -\Phi_X)}}\right]  \,,
\end{equation}
and correspond to rotating the great circle around the sphere, while keeping $\Phi_Y$ and $\Phi_X$ fixed. Note that you can only do this when $\Phi_Y - \Phi_X = \pi$. The length of all these geodesics is the same and reads
\begin{equation}
\tilde{L}_g= \pi \,,
\end{equation}
independently of the choice of $c_1$. In what follows, we will assume that $\Phi_Y - \Phi_X < \pi$.

\subsection{Quadratic perturbations} \label{sec:pert}

We have found geodesics that are saddle points of the Euclidean propagator (\ref{euc_path_int}). Now we can compute the corrections to the propagator stemming from quadratic perturbations to the geodesic length on the sphere. One-loop path integrals on the sphere have been computed in, for instance, \cite{Anninos:2020hfj, Law:2020cpj}. Recall that we are parameterising our paths as $\Theta(\Phi)$, so we want to consider perturbations to the geodesics of the form
\begin{equation}
\Theta (\Phi) = \Theta_{\text{geodesic}} (\Phi) + \delta \Theta (\Phi) \,,
\end{equation}
where the geodesic equation just gives $
\Theta_{\text{geodesic}} (\Phi) = 0 \,. $  The variation of the Euclidean two-point function (\ref{euc_path_int}) is given by\footnote{In principle, the measure in this path integral should include a factor of $\cos(\Theta)$ that comes from the determinant of the metric \eqref{new_metric}. The inclusion of this factor is needed for the path integral to be diffeomorphism invariant \cite{PhysRevD.23.2850}. Note, however, that this term will not contribute to the path integral in the large mass expansion.}
\begin{equation}
\tilde{\mathcal{G}} (\Phi_X, \Phi_Y) \approx  \sum_{*=g,+} \left( e^{-m \tilde{L}_* (\Phi_X, \Phi_Y) }\int D \delta \Theta (\Phi) \exp \left( - m\, \delta\tilde{L}_* (\Phi_X, \Phi_Y, \delta\Theta, \delta\dot{\Theta}) \right) \right) \,, \label{GE}
\end{equation}
where, evaluating the length functional (\ref{length_funct_2}) to second order around each geodesic, we obtain that,
\begin{eqnarray}
\tilde{L} (\Phi_X,\Phi_Y, \delta \Theta)  =  \tilde{L}_{*} (\Phi_X, \Phi_Y) + \delta \tilde{L}_{*} (\Phi_X, \Phi_Y, \delta\Theta, \delta\dot{\Theta})  \,, 
\end{eqnarray}
with
\begin{eqnarray}
\delta \tilde{L}_{*} (\Phi_X, \Phi_Y, \delta\Theta, \delta\dot{\Theta})  \equiv  \frac{1}{2} \int d\Phi \left(\delta \dot{ \Theta} (\Phi)^2 - \delta \Theta (\Phi)^2 \right)\,,
\end{eqnarray}
where the integration limits in the last integral depend on which geodesic we are expanding around, and are the same as in equations  (\ref{L-}) and (\ref{L+}). Given that $\delta\tilde{L}_{*}$ is quadratic in $\delta \Theta$, this path integral can be computed exactly. In general, consider the following quantum mechanical path integral,
\begin{equation}
Z (\Phi_0, \Phi_N) \equiv \int_{\delta \Theta(\Phi_0) = 0}^{\delta \Theta(\Phi_N)=0   } D\delta\Theta (\Phi)  \exp \left( - \frac{m}{2} \int_{\Phi_0  }^{ \Phi_N} d\Phi \left( \delta \dot{ \Theta}^2 - \delta\Theta^2 \right) \right) \,,
\end{equation}
where the generic endpoints of the path integral are named $\Phi_0$ and $\Phi_N$, we assume $\Phi_N > \Phi_0$, and we require a vanishing variation of the trajectory at these points. In Appendix \ref{path_integral_app}, we show how to compute a more general class of quadratic path integrals in quantum mechanics, including this one. The final result is
\begin{equation} \label{one_loop_path_integral}
Z (\Phi_0, \Phi_N)= \sqrt{\frac{m}{2\pi \sin (\Phi_N - \Phi_0)}} \,.
\end{equation}
Inserting this back into equation (\ref{GE}), we can write the correlator between any two points on the sphere, in the large mass limit, as a function of the geodesic length $\tilde{L}_g$ between the two points. This yields, 
\begin{equation} \label{euclidean_one_loop}
\tilde{\mathcal{G}} (\tilde{L}_g) = \sqrt{\frac{m}{2\pi \sin \tilde{L}_g}} e^{-m \tilde{L}_g} + \sqrt{\frac{m}{2\pi \sin(2\pi-\tilde{L}_g)}} e^{-m (2\pi - \tilde{L}_g)} \,,
\end{equation}
which looks suggestively similar to (\ref{GofL}) with $d=2$. We stress that this result is valid for any arbitrary two points on the sphere, as long as $\tilde{L}_g \gtrsim m^{-1}$ and $\pi - \tilde{L}_g \gtrsim m^{-1}$.

As previously mentioned, in Euclidean signature the second term in (\ref{euclidean_one_loop}) will always be exponentially suppressed in the large mass limit. However, we will keep both saddle points, because, interestingly, in some cases, after doing the analytic continuation back to Lorentzian signature, they will both contribute to the Lorentzian, large mass two-point correlator.

\subsection{Higher dimensions} \label{sec_higher_d}

It is possible to generalise the calculation on $S^2$ to higher dimensions. In this case, the analytic continuation of the global dS metric in equation (\ref{global_dS}) is given by the round metric on $S^d$,
\begin{equation} \label{metricSd}
ds^2  = d\theta^2 + \cos^2 \theta \, d\Omega_{d-1}^2 \,.
\end{equation}
In any dimension, it is also true that the geodesics between any two points are sections of the great circle between those two points. As in the case of two dimensions, it is convenient to rotate the coordinates to a frame where the $\Phi$ coordinate goes around the great circle between the two endpoints. We will call these coordinates $\{\Theta_1, \cdots, \Theta_{d-1}, \Phi\}$. In this frame, the metric on $S^d$ is given by
\begin{equation}
ds^2  = d\Theta_1^2 + \cos^2 \Theta_1 d\Theta_2^2 + \cos^2 \Theta_1 \cos^2 \Theta_2 d\Theta^2_3 + \cdots + \cos^2 \Theta_1 \cos^2 \Theta_2 \cdots \cos^2 \Theta_{d-1} d\Phi^2  \,,
\end{equation}
so that the great circle lies at $\Theta_i = 0$. 
As in the two dimensional case, we can use $\Phi$ to parameterise the geodesic, which will follow a path $(\Theta_1(\Phi), \cdots, \Theta_{d-1} (\Phi))$, that extremises the length functional,
\begin{equation} 
		\tilde{L} = \int d\Phi \sqrt{\dot{\Theta}^2_1 + \cos^2 \Theta_1 \left( \dot{\Theta}_2^2 + \cos^2 \Theta_2 \dot{\Theta}_3^2+   \cdots + \cos^2 \Theta_2 \cdots \cos^2 \Theta_{d-1} \right)} \, . 
\end{equation}
In Appendix \ref{app_ddim}, it is shown that the equations of motion imply that the saddle point is given by
\begin{equation}
\Theta_i^{(\text{geodesic})} = 0  \, , \qquad \text{for} \,\, 1 \leq i \leq d-1 \,.
\end{equation}
As in the case of two dimensions, this implies that there will be two geodesics leading the saddle point approximation. The one with minimal length is
\begin{equation} \label{L-d}
\tilde{L}_{g} = \int^{\Phi_Y}_{\Phi_X} d\Phi  = \Phi_Y - \Phi_X \,,
\end{equation}
where again we assume that $0 < \Phi_Y - \Phi_X < \pi$. The other geodesic goes around the remainder of the great circle and has length
\begin{equation} \label{L+d}
\tilde{L}_{+} = \int^{2\pi + \Phi_X}_{\Phi_Y} d\Phi  = 2\pi - (\Phi_Y - \Phi_X) = 2\pi - \tilde{L}_g \,.
\end{equation}
The contributions from other geodesics that wrap around the great circle multiple times will be exponentially suppressed in the two-point function, so we neglect them. We can expand the length functional around each geodesic trajectory, $\Theta_{i} = 0 + \delta \Theta_i$, and this gives 
\begin{equation}
\tilde{L} = \int d\Phi \left(1 + \frac{1}{2} \sum_{i=1}^{d-1} \left( \delta \dot{\Theta}_i^2 - \delta\Theta_i^2 \right)  \right) \,,
\end{equation}
where each of the $(d-1)$ terms in the sum give the same contribution to the path integral, and this is exactly the same contribution as in the two-dimensional case. So, finally, we get that for most\footnote{The same restrictions as in the $d=2$ case apply, \ie $\tilde{L}_g \gtrsim m^{-1}$ and $\pi - \tilde{L}_g \gtrsim m^{-1}$.} two points on the higher dimensional sphere, the Euclidean correlator in the large mass limit can be written as a function of the geodesic length between the two points. The Euclidean correlator, to this order, is given by
\begin{equation} \label{euclidean_higher_d}
\tilde{\mathcal{G}} (\tilde{L}_g) = \left(\frac{m}{2\pi \sin \tilde{L}_g}\right)^{\frac{d-1}{2}} e^{-m \tilde{L}_g} + \left(\frac{m}{2\pi \sin (2\pi  - \tilde{L}_g)}\right)^{\frac{d-1}{2}} e^{-m (2\pi - \tilde{L}_g)} \,.
\end{equation}

%% file: sections/lorentz2pt.tex
We will now use the results of the last section to reproduce the Lorentzian two-point function for a free massive scalar field in the Euclidean state $|E\rangle$. For simplicity, consider $d=2$. On the Euclidean sphere, the geodesic distance between any two points $(\theta_1, \varphi_1)$ and $(\theta_2, \varphi_2)$ is given by 
\begin{equation} \label{spheregeodesicdist}
	\tilde{L}_g = \cos^{-1} \left[\cos \theta_1 \cos \theta_2 \cos (\varphi_2 - \varphi_1) + \sin \theta_1 \sin \theta_2 \right] \, . 
\end{equation}
Analytically continuing back to Lorentzian dS space by taking $\theta \to i T$, we recover $L_g = \cos^{-1} P_{X,Y}$, with $P_{X,Y}$ as in equation \eqref{lenTphi}, even in the regime where Lorentzian geodesics do not exist. So it is straightforward to verify that

\begin{equation} \label{lorentzian_euclidean}
	\left. \tilde{\mathcal{G}} (\tilde{L}_g) \right|_{\theta \to i T} \, = 2 m \, G(L_g) \, . 
\end{equation}
Using (\ref{GofL}) and (\ref{euclidean_higher_d}), one can also verify that the same formula holds in higher $d$.

The apparent tension in section \ref{sec:tension} is now resolved. For $P_{X,Y}<-1$, the complex nature of the Lorentzian geodesic length comes from analytic continuation of Euclidean geodesic lengths on the sphere. In this particular case, the Lorentzian geodesic length can be written as $L_g = \pi - i \cosh^{-1} |P_{X,Y}|$. The second saddle (corresponding to the Euclidean geodesic encircling the sphere from the opposite side) has length $2\pi - L_g$ and so, in Lorentzian signature, they both have the same real part. Thus, neither of them can be neglected in the large mass limit. This explains the need for both terms in (\ref{GofL}).

When $P_{X,Y}>-1$, it is always true that the second saddle is exponentially suppressed, as it will always have a larger real part than $L_g$, after analytically continuing back the Euclidean answer. 

For any $P_{X,Y}$, it is important to keep the next order correction to the saddle point answer in order to reproduce the large mass correlator.

In the remainder of this section, we explore several choices of points that illustrate the different features of the correlator and the geodesics in the different regimes.

\subsection*{Timelike separated points}
As a first example, consider a fixed point on the spatial $S^{d-1}$, at two different global times $T_1$ and $T_2$. In this case, $P_{X,Y} = \cosh (T_2 - T_1)>1$ and the relevant geodesics have a complex length $L_g = i \cosh^{-1} |P_{X,Y}|$.

It follows from (\ref{P_greater}) that in the large mass limit the correlator is given by
\begin{equation} \label{timelike_d}
	G (P_{X,Y})  \approx 
	\frac{m^{\frac{d-3}{2}}}{2^{\frac{d+1}{2}}  \pi^{\frac{d-1}{2}}  \sinh^{\frac{d-1}{2}} \left| T_2-T_1 \right|} e^{-i m \left| T_2-T_1 \right|} e^{-i \frac{\pi}{4} (d-1)} \, . 
	\end{equation}

This correlator can be obtained from the geodesic approximation both in Lorentzian and in Euclidean signature. Given that $L_g$ is purely imaginary, it is clear that the contribution coming from the geodesic with length $\tilde{L}_+ = 2\pi - \tilde{L}_g$  will always have a larger real part when taken to Lorentzian signature, so it will be exponentially suppressed. 

\begin{figure}[H]
        \centering
         \subfigure[Timelike separated points]{
                \includegraphics[height=7cm]{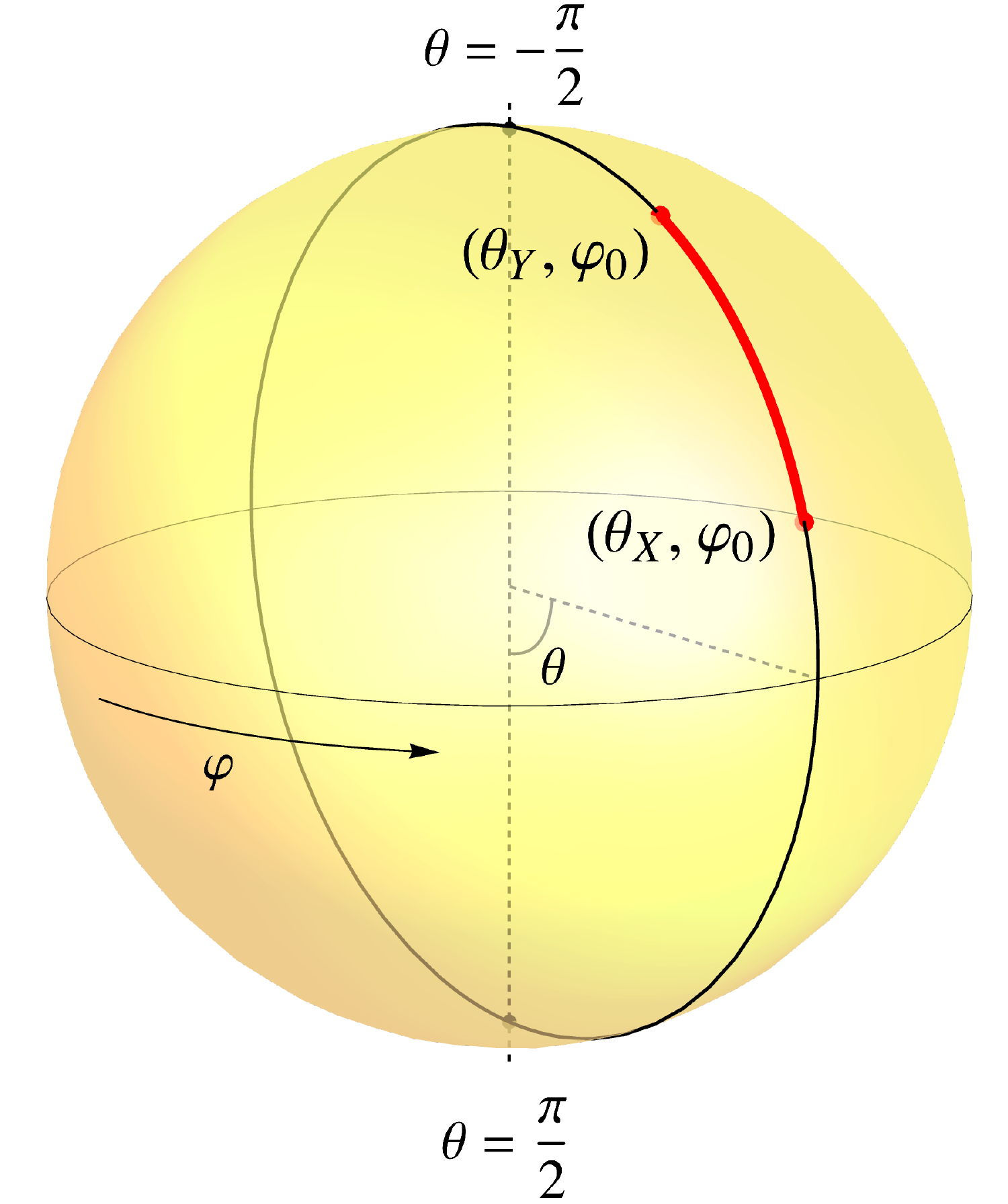}\label{fig:Timelike}}  \qquad \qquad 
        \subfigure[Spacelike separated points]{
                \includegraphics[height=7cm]{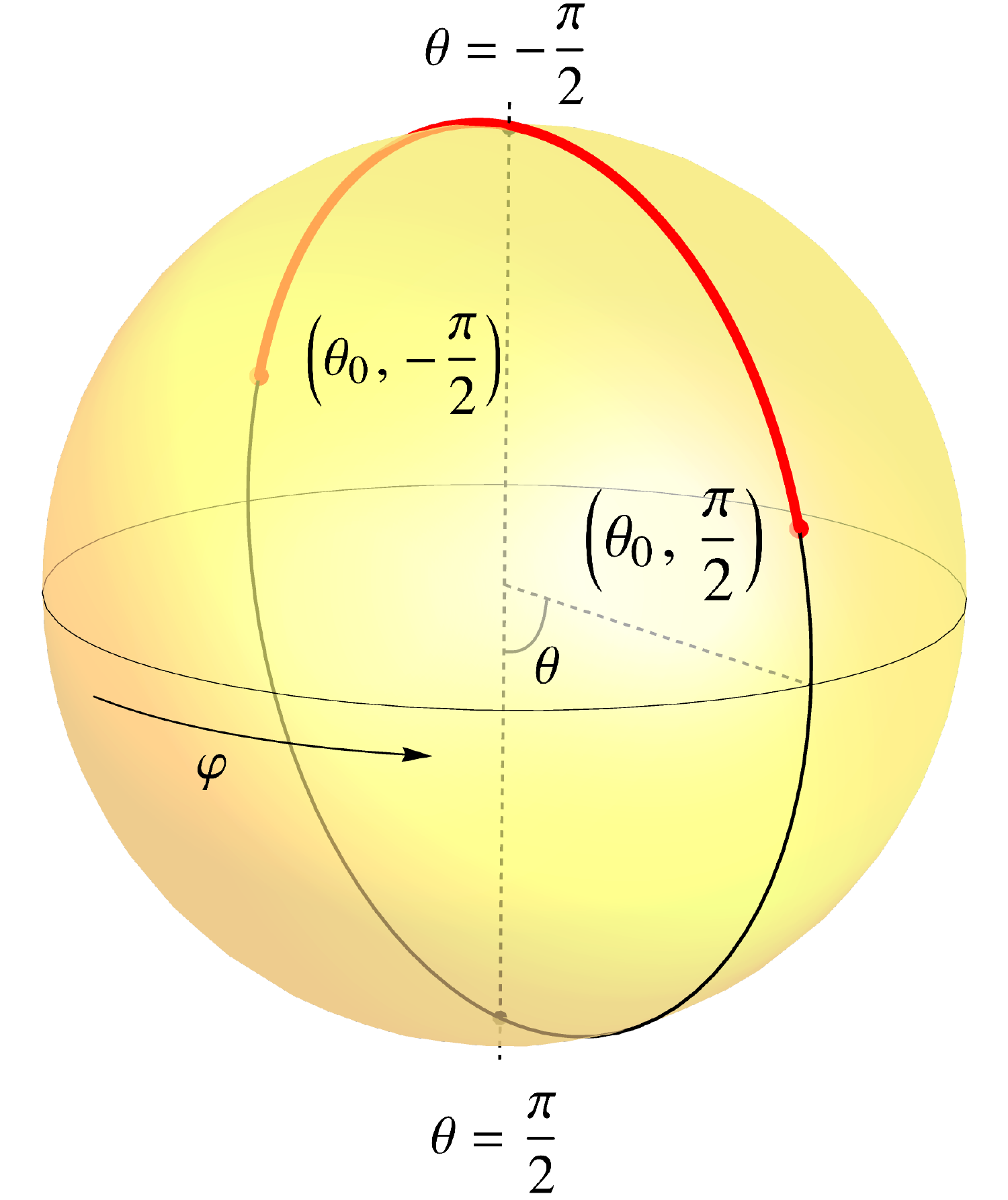} \label{fig:Spacelike}}
                                \caption{Upon analytic continuation to the Euclidean sphere, we look for two different types of geodesics, depending on the type of correlator under consideration. The geodesic with shorter length is shown in red, while the one going on the other side of the great circle is shown in black. }\label{fig:spheres}
\end{figure}

To illustrate this case, consider the sphere for $d=2$. We choose points with the same spatial angle $\varphi = \varphi_0$. On the sphere they will look  as in figure \ref{fig:Timelike}. The Euclidean geodesic length will be given by $\theta_Y - \theta_X$, with $\theta_Y > \theta_X$, and this is enough to reproduce (\ref{timelike_d}) for $d=2$.

\subsection*{Spacelike separated points}

Next, consider opposite points on the $S^{d-1}$ at a given global time $T$. 
In this case, $P_{X,Y} = -\cosh 2T < -1$. 
In the large mass limit, it follows from equation (\ref{Pminus1}) that,
\begin{equation}
G (P_{X,Y})  \approx \frac{m^{\frac{d-3}{2}}}{2^{\frac{d}{2}}  \pi ^{\frac{d-1}{2}}} \frac{ e^{-m \pi}}{\sinh^{\frac{d-1}{2}}2 |T|} \left(\cos \left(2 m |T| - \frac{\pi  d}{4}\right) - \sin \left(2 m |T| - \frac{\pi  d}{4}\right)\right) \,. \label{two_point_d}
\end{equation}
The correlator is real, but it oscillates with a frequency of $2m$. Furthermore, it exponentially decays as a function of time and the decay rate does not depend on the mass. There do not exist real geodesics to account for this behaviour. This result for the two point function was first found for $d=2$ and $d=4$ in \cite{Galante:2022nhj} and \cite{Anninos:2022ujl}, respectively.

Again, to illustrate this behaviour we focus on $d=2$. On the sphere, it is natural to choose points opposite to each other at a given latitude $\theta = \theta_0$ (one at $\varphi = -\tfrac{\pi}{2}$ and the other one at $\varphi = \tfrac{\pi}{2}$), as in figure \ref{fig:Spacelike}. Then, the Euclidean geodesic lengths are given by
\begin{equation} \label{l- spacelike}
\tilde{L}_g (\theta_0) = \pi - 2 \theta_0  \,, \qquad \tilde{L}_+ (\theta_0)  = \pi + 2 \theta_0 \,.
\end{equation}
Note that, in this case, both will have the same real part when we analytically continue back to Lorentzian spacetime, and so neither can be neglected. Plugging this $\tilde{L}_g$ into (\ref{euclidean_one_loop}), it is straightforward to verify (\ref{lorentzian_euclidean}).

\subsection*{Spacelike separated points between stretched horizons}
So far we have only considered points for which either $P_{X,Y} >1 $ or $P_{X,Y}  < -1$. It is interesting to consider a case where $P_{X,Y}$ goes through the transition point $P_{X,Y} = -1$. 

\begin{figure}[H]
        \centering 
        \subfigure[Penrose diagram]{
                \includegraphics[height=5.85cm]{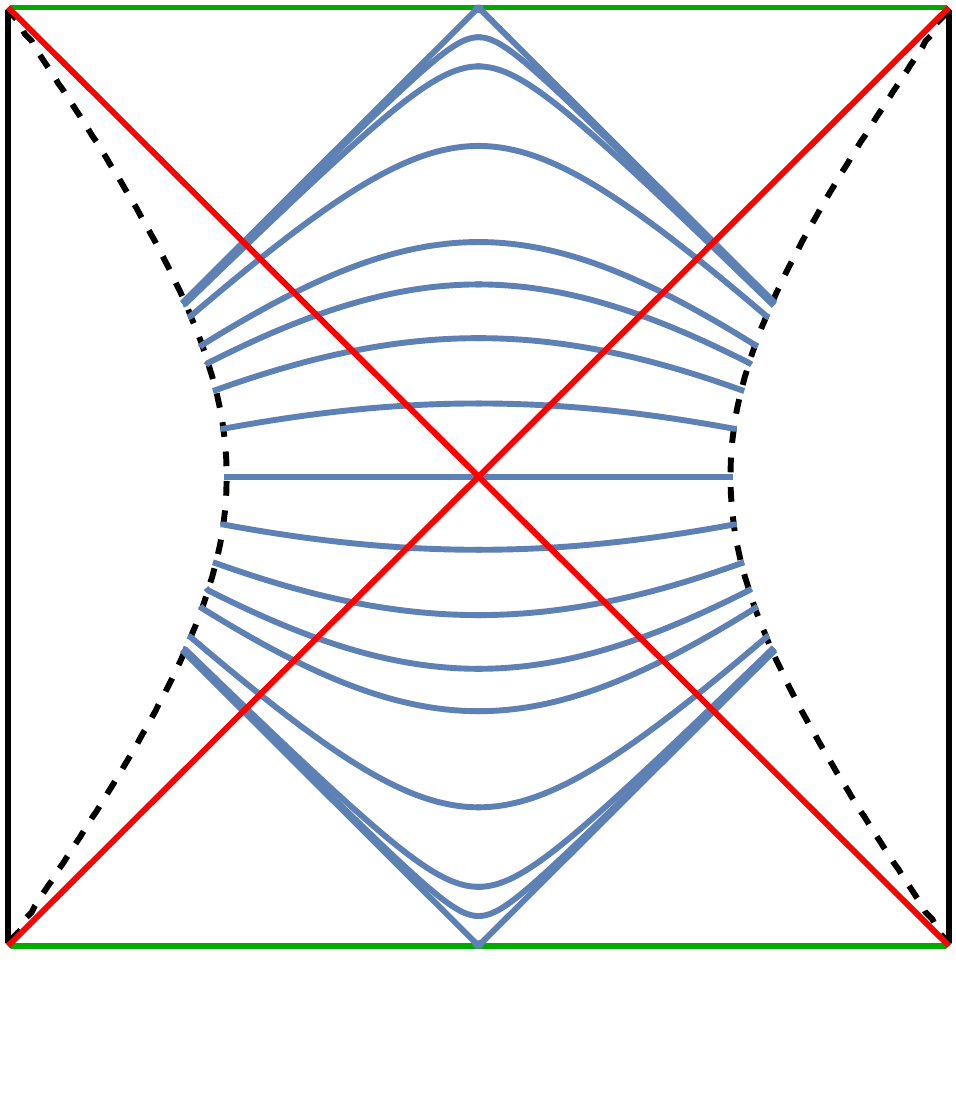} \label{fig:stretched}} \qquad
                  \subfigure[Correlator]{
                \includegraphics[height=5.8cm]{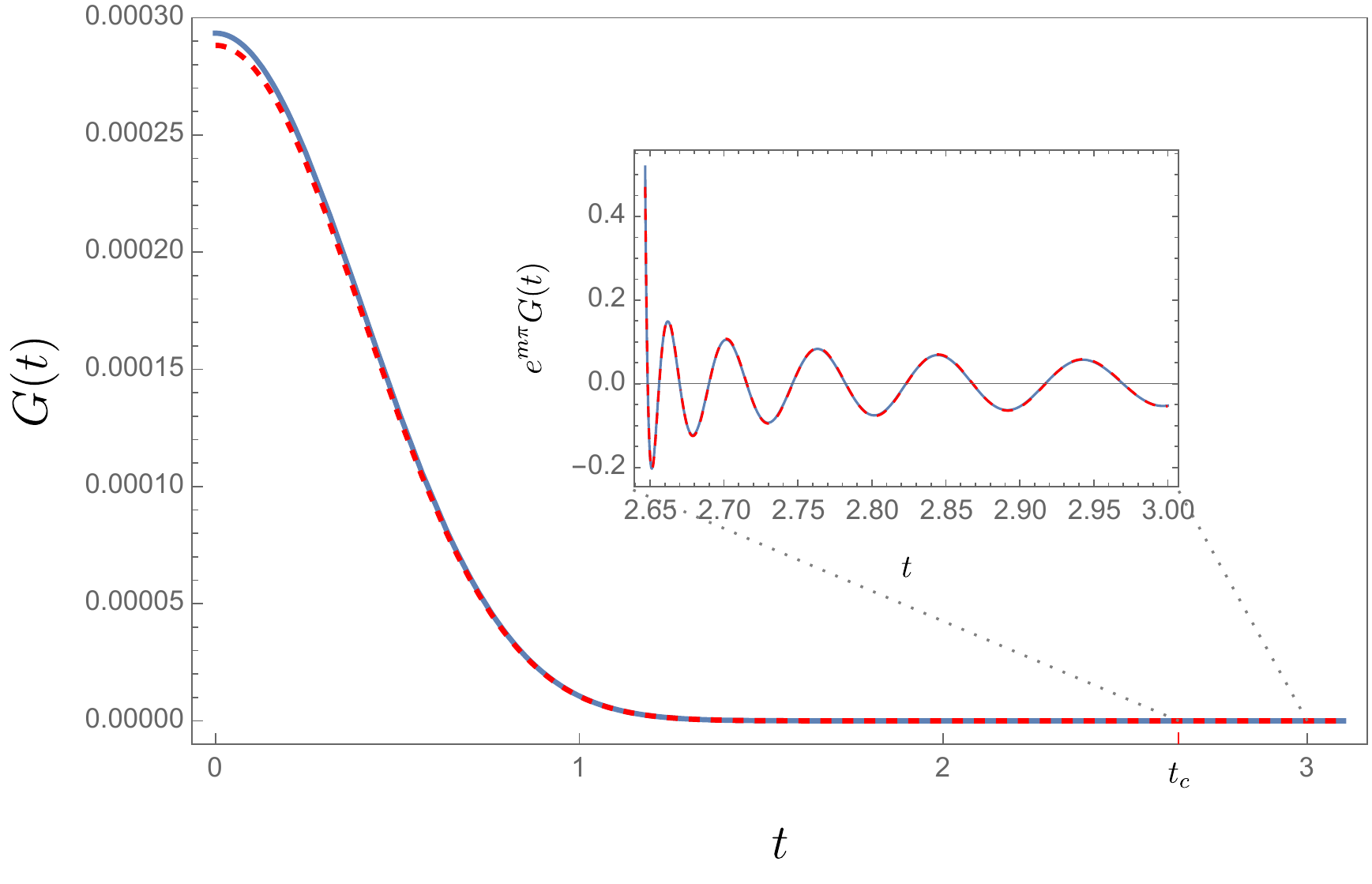}\label{fig:stretchedP}}  
                               
\caption{(a) Penrose diagram showing the geodesics (in blue) between two stretched horizons (in dashed black). The position of the stretched horizon is set to $r_{st}=2/3$. (b) Exact (dashed red) and large mass (solid blue) correlator for two symmetric points on opposite stretched horizons as a function of the static patch time. At early times  we observe a decay of the correlation function while at later times we observe an oscillatory behaviour. We multiply by $e^{m\pi}$ in the inset to make the oscillations apparent. In the plot, $m=20$ and $r_{st}=0.99$, so $t_c \sim 2.65$. Given the stretched horizon is very close to the actual horizon, at $t\sim0$, the two points become very close to each other, and so the approximation breaks down.}
\end{figure}

A concrete example of this involves studying the form of the correlator  between points anchored at opposite stretched horizons, as a function of the static time; see figure \ref{fig:stretched}.  For simplicity, we again restrict to $d=2$. As discussed, all we need in order to find the correlator is $P_{X,Y}$ between these points. The stretched horizon  $r_{st}$ is defined as a constant $r$ surface in the static patch metric (\ref{static_metric}). In order to get the position of the opposite stretched horizon, we need to relate static patch coordinates to global coordinates. Explicit expressions are shown in Appendix \ref{app:details}.

It is clear that opposite points on the Penrose diagram with a fixed $r = r_{st}$ will have the same $T_0$ and their angles will be at $\varphi_0$ and $- \varphi_0$. 
In this setup then, one point is at embedding coordinates $X$ and the second one, $Y$, is at the same coordinates except for $Y_2 = - X_2$. This gives,
\begin{equation} 
    P_{X,Y}^{st} = -(X_0)^2+(X_1)^2-(X_2)^2 = \left(r_{st}^2-1\right) \cosh 2 t+r_{st}^2 \,. \label{StretchedP}
\end{equation} 
Note that $P_{X,Y}^{st} = -1$ at the critical time
\begin{equation}
t_c \equiv \frac{1}{2} \cosh ^{-1}\left(\frac{1+r_{st}^2}{1-r_{st}^2}\right) = \frac{1}{2} \log \left(\frac{1+r_{st}}{1-r_{st}}\right) \,. \label{critical_t}
\end{equation} 
For times between $0 < |t| < t_c$, the de Sitter invariant length has range $1 > P_{X, Y}^{st} > -1$, but for $|t| > t_c$, the range becomes $P_{X,Y}^{st} < -1$. The large mass correlator will then be given by (\ref{P_greater}) with $P_{X,Y}^{st}$ before $t_c$, and by (\ref{Pminus1}), after. We plot the correlator in figure \ref{fig:stretchedP}. Before $t_c$ the correlator decays monotonically, while afterwards it starts oscillating with frequency proportional to the mass.

We can easily obtain geodesics anchored at the stretched horizon by simply cutting parts of the geodesics obtained in section \ref{sec_geo}. See figure \ref{fig:stretched}. Note that this breaks the degeneracy of geodesics at $T=0$, and now there is at most one geodesic at each static time $t$. Their length is simply given by $L_g = \cos^{-1} P_{X,Y}^{st}$, up to $|t| = t_c$. At these times, the last geodesics have length $L_g = \pi$, as we are just removing two null pieces from the geodesics from section \ref{sec_geo} that have that same length.

After $t_c$, there are no more real geodesics. To find the complex geodesics, on the sphere we look for points at $(\theta_0,\varphi_0)$ and $(\theta_0, - \varphi_0)$. Using the geodesic length formula (\ref{spheregeodesicdist}), we find that
\begin{equation}
	\tilde{L}_g =  \cos^{-1} \left[\cos^2 \theta_0 \cos ( 2\varphi_0) + \sin^2 \theta_0 \right] \,  , 
\end{equation}
and for the longer geodesic, we have $\tilde{L}_+ = 2 \pi - \tilde{L}_g $. After analytic continuation, we can transform the global coordinates into static patch ones, and this Euclidean geodesic length recovers the Lorentzian two-point function, as in (\ref{lorentzian_euclidean}). For $|t|>t_c$, we do need the contributions from both geodesics. But it is interesting to note that the Euclidean length above works for both $P_{X,Y}^{st}$ smaller and larger than $-1$, so it seems that the Euclidean computation does not know about $t_c$.

%% file: sections/outlook.tex
In this paper, we studied the two-point correlator of a free massive scalar field in a fixed dS background, in the Euclidean state. The aim of this work was to reproduce the asymptotic form of the correlator in the large mass limit using the geodesic approximation. This was naively puzzling, since certain points in dS are not connected by geodesics. The resolution was to look at the Euclidean problem, where on the sphere any two points are connected around a great circle by two geodesics. Upon analytically continuing them back to Lorentzian space, we found that these geodesic can have complex lengths, that they both contribute to the correlator in the large mass limit and that, complemented with the one-loop correction around each geodesic length, they give the precise asymptotic form of the two-point correlator (up to a proportionality factor). This result is valid for almost any two points in any number of dimensions.

We then studied the correlator for some particular choices of points, including both timelike and spacelike correlators, and we  interpreted them in terms of Euclidean geodesics. Given the perfect matching we found, one can ask whether this prescription can be used in more general setups (of course, in this case, the choice of a Euclidean state was crucial to obtain the right results).

In particular, the geodesic approximation has been widely used in the context of holography, starting with the work in \cite{Balasubramanian:1999zv, Louko:2000tp}, where boundary conformal correlators of heavy operators were reproduced both at zero and finite temperature from a bulk geodesic calculation. Geodesics have also been used to explore dynamical settings in holography, such as in \cite{Balasubramanian:2011ur, Aparicio:2011zy, Liu:2013iza}, and quantum chaos \cite{Shenker:2013pqa, Shenker:2013yza}, among other things. Complex geodesics in AdS have been studied, for instance, in \cite{Fidkowski:2003nf, Festuccia:2005pi, Balasubramanian:2012tu}.

Recently, there has been an increased interest in applying the standard tools of AdS/CFT to probe the static patch of dS. A non-comprehensive list includes \cite{Banks:2003cg, Banks:2004eb, Banks:2006rx, Parikh:2004wh, Banks:2018ypk, Geng:2019ruz,  Aalsma:2020aib, Geng:2020kxh, Aalsma:2021bit, Shyam:2021ciy, Coleman:2021nor,  Svesko:2022txo, Banihashemi:2022jys, Silverstein:2022dfj, Nomura:2017fyh, Nomura:2019qps, Murdia:2022giv}. One approach is to study flow geometries that interpolate between an AdS boundary and a dS interior \cite{Anninos:2017hhn, Anninos:2018svg, Anninos:2022hqo}.\footnote{Most of these constructions are two dimensional, but see \cite{Anninos:2022ujl} for higher dimensional examples.} Here, the presence of an asymptotic, timelike boundary permits us to interpret bulk correlators in terms of correlation functions of the boundary quantum mechanics. For heavy fields, we can employ a geodesic approximation to compute such correlators. In the two sided geometries, the result is that geodesics between opposite boundaries only exist for a short period of time (of order of the inverse temperature), after which there are no more real geodesics \cite{Chapman:2021eyy}. The last geodesic is almost null and goes all the way to the future (or past) infinity. This is reminiscent of what happens for the AdS double-sided black hole in dimensions higher than 3. In that case, the singularity in the Penrose diagram bends inwards, and spacelike geodesics from the boundary are able to reach it \cite{Fidkowski:2003nf, Festuccia:2005pi}. However, the boundary correlator is not expected to have singularities of this type. The resolution is that there exist complex geodesics, whose contributions to the correlator are necessary to reproduce the correct answer for the boundary correlator. A similar story might hold for flow geometries where both the correlator and the geodesics can be computed in the bulk. 

From a different perspective, it has also been advocated that the dual theory to dS might live on a stretched horizon \cite{Susskind:2021dfc, Shaghoulian:2021cef, Shaghoulian:2022fop}. The microscopic candidate theory is an SYK model in a particular double-scaled limit \cite{Susskind:2021esx, Lin:2022nss, Susskind:2022bia, Rahman:2022jsf}. We computed correlation functions of bulk scalar fields anchored at opposite stretched horizons in section \ref{lorentzian_props}. The conclusion is that real geodesics exist only up to some critical time $t_c$, that depends on the position of the stretched horizon. After this time, the correlator starts exhibiting oscillations, and the geodesic length becomes complex. Using holography, we would expect the boundary theory to know about this time scale. It looks hard to envision how a standard SYK model would incorporate this scale. Other proposals relating non-Hermitian SYK models to dS include \cite{Anninos:2020cwo, Garcia-Garcia:2022adg, sam}. 

\begin{figure}[H]
        \centering
        \includegraphics[height=4cm]{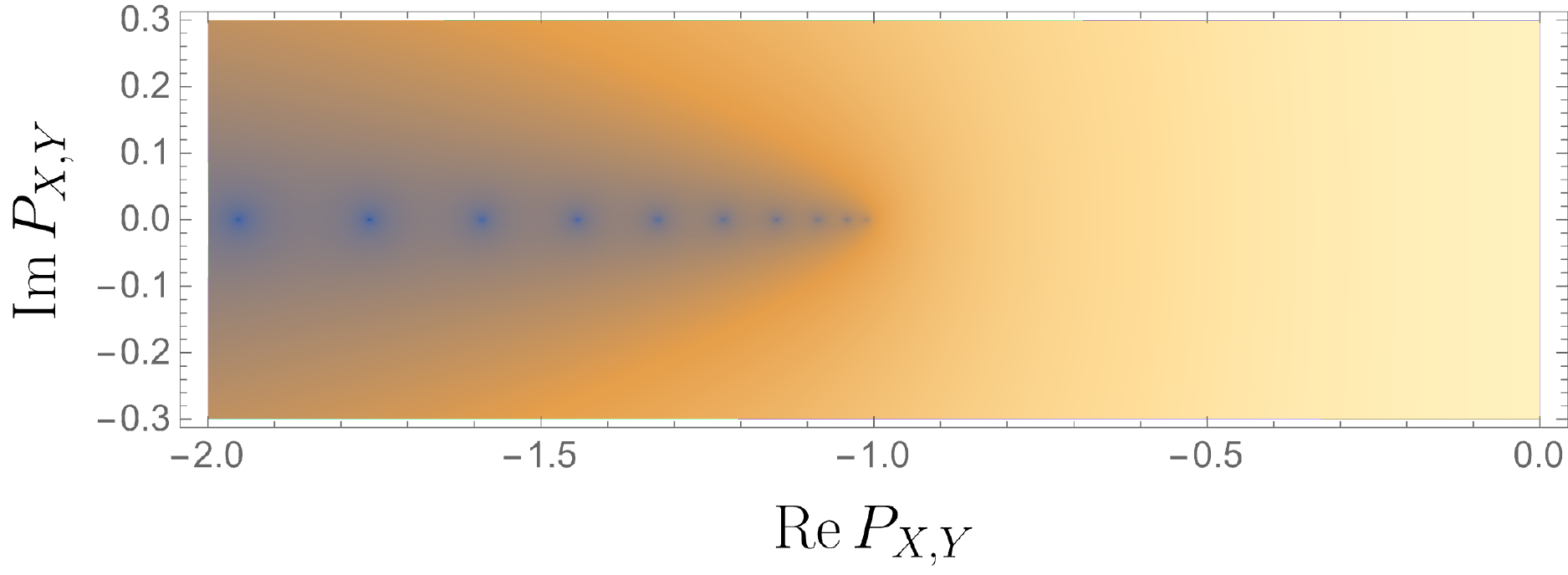}
\caption{Density plot of $|G(P_{X,Y})|$ for $d=4$ and $m=25$. In the large-$m$ limit zeroes (blue dots) accumulate near $P_{X,Y}=-1$. }\label{fig:zeroes}
\end{figure}

One might worry that some of the effects shown here might be hard to observe, due to the exponential suppression in the large mass limit. However, the characteristic oscillations\footnote{As in the case of the AdS black hole \cite{Fidkowski:2003nf, Amado:2008hw}, these oscillations are probably related to the quasinormal mode frequencies of the cosmological horizon \cite{Lopez-Ortega:2006aal}.} of the two-point function will already be present for large separations in the exact correlator as soon as  $m>(d-1)/2$. A harder question is whether there are signatures of the breakdown of the approximation in the exact correlator at $P_{X,Y}=-1$ in the large mass limit. As in \cite{Fidkowski:2003nf}, we see \textit{subtle but distinct} signatures of this breakdown in the exact correlator that show up as an accumulation of zeroes near $P_{X,Y}=-1$ as we increase the mass; see figure \ref{fig:zeroes}.

It would be interesting to study the role of complex geodesics in interacting quantum field theories in dS, perhaps along the lines of the K\"{a}llen-Lehmann representation \cite{Bros:1990cu}, or in other backgrounds with positive cosmological constant, such as dS black holes.

Finally, geodesics are not the only interesting extended extremal objects used  in holography. For instance, co-dimension 2 surfaces are related to entanglement entropy \cite{rangamani2017holographic}, and co-dimension 1 (or zero), to holographic complexity
\cite{Susskind:2018pmk,Chapman:2021jbh}. It is reasonable to expect the existence of complex surfaces in the context of dS \cite{Fischetti:2014uxa}; see also \cite{Jorstad:2022mls}. However, it would be hard to interpret complex areas or volumes as measures of these naturally real quantities. As we have seen, in the case of the two-point function, the complex geodesics combine such that the final answer is real. It would be interesting to understand the more general role that complex surfaces might play in holography.

%% file: sections/appwkb.tex
The Klein-Gordon equation as a function of $P\equiv P_{X,Y}$ can be solved systematically, order by order in the large mass expansion, through a WKB approach. A similar procedure is followed in Appendix B of \cite{Balasubramanian:2019stt} for the case of timelike geodesics in AdS.

To start, we recall here equation (\ref{kg_P}),
\begin{equation}
\label{eq:exa}
(1-P^2) \partial_{P}^2 G(P) - d P \partial_{P} G(P) - m^2 G(P) = 0 \,.
\end{equation}
We take the ansatz,
\begin{equation}
G(P) \propto \exp \left( -m X(P)  \right) \,,
\end{equation}
where $X$ is assumed to be independent of $m$. At leading order in the large mass limit, it satisfies the equation,
\begin{equation}
\left(P^2-1\right) X'(P)^2+1=0 \,.
\end{equation}
This equation is solved by
\begin{eqnarray}
\label{eq:wkbsol1}
X_\pm^{(1)}(P) &=& c_\pm^{(1)} \pm \cos^{-1} P \,,  \qquad \textrm{for} \, P<-1 \, \,, \\
\label{eq:wkbsol2}
X_\pm^{(2)}(P) &=& c_\pm^{(2)} \pm \cos^{-1} P \,,  \qquad \textrm{for} \, -1<P<1 \, \,, \\
\label{eq:wkbsol3}
X_\pm^{(3)}(P) &=& c_\pm^{(3)} \pm \cos^{-1} P \,,  \qquad \textrm{for} \,  P>1 \, \,.
\end{eqnarray}
The six constants $c_\pm^{(i)}$  will be fixed later. To next order, we add functions $A_\pm^{(i)}(P)$, such that now,
\begin{equation}
G^{(j)}(P) = m^\alpha  \sum_{i=\pm}A_i^{(j)}(P) \exp \left( -m X_i^{(j)}(P)  \right) \,,
\end{equation}
where $\alpha$ is an arbitrary constant and we again assume that the functions $A_\pm^{(i)}(P)$ are independent of $m$. Plugging this back in the Klein-Gordon equation, we obtain
\begin{equation}
(d-1) P A_i^{(j)} +2 \left(P^2-1\right) \partial_{P} A_i^{(j)}  = 0 \,,
\end{equation}
which is solved by
\begin{equation}
A_i^{(j)}(P) = a_i^{(j)}(1-P^2)^{\frac{1-d}{4}}\, ,
\end{equation}
where $a_i^{(j)}$ are constant factors.  Although $a_i^{(j)}$ and $c_\pm^{(i)}$ are a priori not independent, they represent multiplicative factors with different powers of $m$, and thus they have to be treated separately.

The general solution we obtain for the propagator at this order is given by
\begin{equation}
G^{(j)}(P) = \frac{m^\alpha}{(1-P^2)^{\frac{d-1}{4}}} \left[ a^{(j)}_+ e^{-m  (c^{(j)}_+ + \cos^{-1} P)} +   a^{(j)}_- e^{-m  (c^{(j)}_- - \cos^{-1} P)} \right] \,.
\end{equation}

The solutions (\ref{eq:wkbsol1})-(\ref{eq:wkbsol3}) break down near $P=\pm 1$ when $1-P^2 \sim m^{-2}$. 
In order to connect the solutions in the three disjoint domains, we need to solve (\ref{eq:exa}) exactly near $P=\pm 1$ and then match the solution to the adjacent WKB solutions.

Let us start with $P=-1$. In the vicinity of this point, (\ref{eq:exa}) simplifies to
\begin{equation}
2(P+1) \partial_{P}^2 G(P) + d \partial_{P} G(P) - m^2 G(P) = 0 \,,
\end{equation}
which is solved by
\begin{equation}
  \label{eq:pm1ex}
  G(P) = (P+1)^{{1\over 2}- {d\over 4}} \left[ a I_{{d\over 2}-1}\left(m\sqrt{2P+2}\right)+
  b K_{{d\over 2}-1}\left(m\sqrt{2P+2}\right) \right] \, ,
\end{equation}
where $a$ and $b$ are two constants and $I$ and $K$ are modified Bessel functions. In order to recover (\ref{t_zero_d}), we need to set
\begin{equation}
  a=2^{{1\over 2} - {3d \over 4}} m^{{d\over 2}-1} \pi^{1-{d\over 2}} e^{-m \pi} \, , \qquad b=0 \, .
\end{equation}
This completely fixes the form of the correlator. We now proceed to match this locally exact solution to the WKB solutions in (\ref{eq:wkbsol1})-(\ref{eq:wkbsol2}). For $P<-1$ taking the large-$m$ limit of (\ref{eq:pm1ex}) gives
\begin{equation}
  \label{eq:lm1}
  G(P) = -2^{{3\over 4}(1-d)}m^{d-3 \over 2} \pi^{{1\over 2}-{d\over 2}}  e^{-m\pi} (-1-P)^{{1\over 4}-{d\over 4}} \sin\left( m\sqrt{-2-2P}- {\pi (d+1) \over 4}\right) \,.
\end{equation}
Using $\cos^{-1} P \approx \pi - i \sqrt{2}\sqrt{-1-P}$, 
(\ref{eq:wkbsol1}) can be matched to (\ref{eq:lm1}). This fixes the constants to be
\begin{equation}
  \nonumber
    \alpha = {d-3 \over 2} \, , \quad a_-^{(1)} = i^{d-1} 2^{-{d+1 \over 2}} \pi^{1-d \over 2} \, , \quad  a_+^{(1)} = 2^{-{d+1 \over 2}} \pi^{1-d \over 2}  \, , \quad c_-^{(1)}= 2\pi\, , \quad c_+^{(1)} = 0  \, .
\end{equation}
Plugging these back into (\ref{eq:wkbsol1}), we obtain the simple expression
\begin{equation}
\boxed{
\qquad
  G^{(1)}(P) = m^{d-3 \over 2}  (2\pi)^{1-d \over 2} \textrm{Re}\left[ {e^{-m \cos^{-1} P }\over (1-P^2)^{d-1\over 4}} \right]  \, , \quad \textrm{valid for} \ P<-1 \,.
\qquad
  }
\end{equation}
This expression recovers (\ref{Pminus1}), which has been obtained from a large-$m$ limit of the exact (hypergeometric) solution to (\ref{eq:exa}).

Similarly, for $P>-1$ taking the large-$m$ limit of (\ref{eq:pm1ex}) gives
\begin{equation}
  \label{eq:lm2}
  G(P) =  2^{-{3d+1\over 4}}   \pi^{{1\over 2}-{d\over 2}} (1+P)^{{1\over 4}-{d\over 4}} e^{m\sqrt{2+2P}-m\pi} \, .
\end{equation}

Using $\cos^{-1} P \approx \pi - \sqrt{2}\sqrt{P+1}$, this can be matched to the solution in (\ref{eq:wkbsol2}) with the plus sign,
\begin{equation}
  \nonumber
  a_-^{(2)}=0 \, , \qquad a_+^{(2)} =    2^{-{d+1 \over 2}} \pi^{1-d \over 2}  \, , \qquad c_+^{(2)} = 0 \, .
\end{equation}
Note that the only difference compared to the $P<-1$ solution is that the $X_-$ component is exponentially suppressed and therefore no longer present.

The WKB solution to the correlator   becomes
\begin{equation}
\boxed{
\qquad  \label{eq:g2sol}
  G^{(2)}(P) = {1\over 2} \times m^{d-3 \over 2}  (2\pi)^{1-d \over 2}  {e^{-m \cos^{-1} P }\over (1-P^2)^{d-1\over 4}} \, , \quad \textrm{valid for} \ -1<P<1  \,. \qquad
  }
\end{equation}
The expression recovers (\ref{P_greater}).

In order to connect the solution in the middle domain to that in $P>1$, we need to investigate the $P\approx 1$ region.
Here we have the locally exact solution
\begin{equation}
  \label{eq:pp1ex}
  G(P) = (1-P)^{{1\over 2}- {d\over 4}} \left[ \tilde a I_{{d\over 2}-1}\left(m\sqrt{2-2P}\right)+
  \tilde b K_{{d\over 2}-1}\left(m\sqrt{2-2P}\right) \right] \, .
\end{equation}
Matching it to (\ref{eq:g2sol}) fixes the constants
\begin{equation}
  \tilde a = 0 \, , \qquad \tilde b = 2^{{1\over 2}- {3d \over 4}} m^{{d\over 2}-1} \pi^{-{d\over 2}} \, .
\end{equation}
\textit{En passant}, we note that matching (\ref{eq:pp1ex}) to (\ref{eq:wkbsol3}) implies that (\ref{eq:g2sol}) remains valid in the $P>1$ region,
\begin{equation}
\boxed{
\qquad  G^{(3)}(P) = G^{(2)}(P)  \, , \quad \textrm{valid for} \  P>1 \,. \qquad
  }
\end{equation}

Finally, note that in this WKB approximation the equation is solved order-by-order in $m$, and so this method validates the subtle limit of the hypergeometric function that we took in the main text.

%% file: sections/apppathintegral.tex
In this appendix, we would like to evaluate the following (Euclidean) path integral, 
\begin{equation}
Z (\Phi_0, \Phi_N) = \int^{\delta \Theta (\Phi _N) = 0}_{\delta \Theta (\Phi_0) = 0} D \delta \Theta \exp \left( - m \int_{\Phi_0}^{\Phi_N} d\Phi L(\Phi, \delta \Theta , \delta \dot{\Theta}) \right) \,,
\end{equation}
for a generic quadratic Lagrangian of the form
\begin{equation}
L(\Phi, \delta \Theta , \delta \dot{\Theta}) = \frac{1}{2} \delta \dot{\Theta}^2 + \frac{\alpha(\Phi)}{2} \delta \Theta^2 \,,
\end{equation}
with $\alpha(\Phi)$ an arbitrary function of $\Phi$. 
This path integral is the quadratic correction to the saddle-point solution, and so we want the fluctuations to vanish at the endpoints. This is a textbook path integral that can be solved by discretising the $\Phi$ interval, see for instance \cite{Schulten2}.

By definition, we would like to compute
\begin{multline}
Z (\Phi_0, \Phi_N) = \lim_{N\to \infty} \left( \frac{m}{2\pi \Delta \Phi} \right)^{N/2} \\
\int d\delta \Theta_1 \cdots \int d\delta \Theta_{N-1} \exp \left( - m \Delta \Phi \sum_{j=0}^{N-1} \left( \frac{ (\delta \Theta_{j+1} - \delta \Theta_j)^2}{2 \Delta \Phi^2}  + \frac{\alpha_j}{2} \delta \Theta_j^2\right) \right) \,,
\end{multline}
where $\Phi_j = \Phi_0 + j \Delta \Phi $ and $ \alpha_j = \alpha (\Phi_j)$. The exponent in the previous equation can be written in a quadratic form,
\begin{equation}
 \left( - m \Delta \Phi \sum_{j=0}^{N-1} \left( \frac{ (\delta \Theta_{j+1} - \delta \Theta_j)^2}{2 \Delta \Phi^2}  + \frac{\alpha_j}{2} \delta \Theta_j^2\right) \right) = - \sum_{j,k=1}^{N-1} \delta \Theta_j \, a_{jk} \, \delta \Theta_k \,,
\end{equation}
where $a_{jk}$ are the matrix elements of the following $(N-1) \times (N-1)$ matrix,
\begin{equation}
\left( a_{jk} \right) =  \frac{m}{2 \Delta\Phi} \begin{bmatrix} 
    2 & -1 & 0 & \dots  & 0 & 0 \\
    -1 & 2 & -1 & \dots & 0 & 0 \\
    0 & -1 & 2 & \dots & 0 & 0 \\
    \vdots & \vdots & \vdots & \ddots & \vdots & \vdots \\
      0 & 0& 0 &\dots  & 2 & -1  \\
    0 & 0& 0 &\dots  & -1 & 2 
    \end{bmatrix}
    + \frac{m \Delta\Phi}{2} 
  \begin{bmatrix} 
    \alpha_1 & 0 & 0 & \dots  & 0 & 0 \\
    0 & \alpha_2 & 0 & \dots & 0 & 0 \\
    0 & 0 & \alpha_3 & \dots & 0 & 0 \\
    \vdots & \vdots & \vdots & \ddots & \vdots & \vdots \\
       0 & 0& 0 &\dots  & \alpha_{N-2} & 0  \\
    0 & 0& 0 &\dots  & 0 & \alpha_{N-1} 
    \end{bmatrix} \,.
\end{equation}
If $\det (a_{jk}) \neq 0$, then we can do the multiple Gaussian integrals to obtain, 
\begin{equation}
Z (\Phi_0,  \Phi_N) = \lim_{N\to \infty} \left( \frac{m}{2\pi \Delta \Phi} \right)^{N/2} \left( \frac{\pi^{N-1}}{\det a_{jk}} \right)^{1/2} = \lim_{N\to \infty} \left( \frac{m }{2 \pi} \frac{1}{\Delta \Phi \left( \frac{2 \Delta \Phi}{m } \right)^{N-1} \det a_{jk} } \right)^{1/2} \,. \label{Ztheta}
\end{equation}
It is convenient to define the following function, 
\begin{equation}
f(\Phi_0,\Phi_N) \equiv \lim_{N \to \infty}  \left(  \Delta \Phi \left( \frac{2 \Delta \Phi}{m } \right)^{N-1} \det a_{jk} \right) \,. \label{def_f}
\end{equation}
In order to take the $N \to \infty$ limit, we first define the discrete determinant,
\begin{equation}
\begin{split}
D_{N-1} &\equiv \left( \frac{2 \Delta \Phi}{m } \right)^{N-1} \det a_{jk} \\
&= \begin{vmatrix} 
    2 + \Delta\Phi^2 \alpha_1 & -1 & 0 & \dots  & 0 & 0 \\
    -1 & 2+ \Delta\Phi^2 \alpha_2 & -1 & \dots & 0 & 0 \\
    0 & -1 & 2 + \Delta\Phi^2 \alpha_3 & \dots & 0 & 0 \\
    \vdots & \vdots & \vdots & \ddots & \vdots & \vdots \\
      0 & 0& 0 &\dots  & 2 + \Delta\Phi^2 \alpha_{N-2}& -1  \\
    0 & 0& 0 &\dots  & -1 & 2 + \Delta\Phi^2 \alpha_{N-1} 
    \end{vmatrix}
\,.
\end{split}
\end{equation}
Now we assume that the dimension of the matrix is variable, so $n = N-1$ can vary. We take the determinant using the last column to obtain the following recursion relation,
\begin{equation} \label{recursionrel}
D_n = \left( 2 + \Delta\Phi^2 \alpha_n \right) D_{n-1} - D_{n-2} \,.
\end{equation}
The first two determinants are given by $D_1 = 2 + \Delta\Phi^2 \alpha_1$ and $D_2 = 3 + 2 \Delta \Phi^2 (\alpha_1 + \alpha_2) + \Delta \Phi^4 \alpha_1 \alpha_2$. Using these along with (\ref{recursionrel}), we can analytically continue the discrete determinant to $n \in \mathbb{Z}_{\leq 0}$, finding, for example, $D_0 = 1$ and $D_{-1} = 0$. Rewriting the recursion relation in a more suggestive way, we obtain
\begin{equation}
\frac{D_{n+1} - 2 D_n + D_{n-1}}{\Delta \Phi^2} = \alpha_{n+1} D_n \,.
\end{equation}
But now we are interested in the continuum limit of this expression, where $\Delta\Phi \to 0$ (or $N \to \infty$), so we may interpret the above recursion relation as a second order differential equation in the variable $\Phi = N \Delta\Phi + \Phi_0$. Note that the extra factor of $\Delta \Phi$ in (\ref{def_f}) cancels since the above equation is linear in $D$. We then get,
\begin{equation}
\frac{d^2 f(\Phi_0, \Phi)}{d\Phi^2} = \alpha (\Phi) f(\Phi_0, \Phi) \,.
\end{equation}
The boundary conditions $D_0 = 1$ and $D_{-1} = 0$ give,
\begin{eqnarray}
f(\Phi_0, \Phi_0) & = & \lim_{N \to \infty} \Delta\Phi D_{-1} = 0 \,, \\
\left. \frac{df (\Phi_0, \Phi)}{d\Phi} \right|_{\Phi = \Phi_0} & = &  \lim_{N \to \infty} \Delta \Phi \frac{(D_0 - D_{-1})}{\Delta\Phi}  = 1 \,. 
\end{eqnarray}
Now that we have $f$, we can go back to (\ref{Ztheta}) to find the final result for the path integral, which reads
\begin{equation}
Z (\Phi_0, \Phi_N) = \sqrt{\frac{m}{2\pi f(\Phi_0, \Phi_N)}} \,.
\end{equation}
In the main text we have $\alpha(\Phi) = -1$, so we get $f(\Phi_0, \Phi_N) = \sin (\Phi_N - \Phi_0)$. This is the result we use in the main text to compute the first correction to the geodesic length, see (\ref{one_loop_path_integral}).

%% file: sections/appd-dimensionaleom.tex
In this appendix, we show that the geodesic equations for the $d$-dimensional sphere are solved by $\Theta_i =0$. The metric of dS$_d$ can be written compactly as  
\begin{equation}
	ds^2 = \sum_{a=1}^d \left( \prod_{m = 1}^{a-1} \cos^2 \Theta_m \right) d \Theta_a^2 \, ,
\end{equation}
where $ -\tfrac{\pi}{2} \leq \Theta_i \leq \tfrac{\pi}{2}$ for $1 \leq i < d - 1$ and $0 \leq \Theta_{d}  < 2 \pi$ with $\Theta_{d}   \equiv \Phi$. 

As in the two-dimensional case, we can use $\Phi$ to parameterise the geodesic so that it will follow a path $(\Theta_1(\Phi), \cdots, \Theta_{d-1} (\Phi))$, that extremises the length functional,
\begin{equation}  \label{ddimlength}
		\tilde{L}^d = \int d\Phi \sqrt{\ \sum_{a=1}^{d} \left( \prod_{m = 1}^{a-1} \cos^2 \Theta_m \right) \dot{\Theta}_a^2 } \, ,
\end{equation}
where again the dot represents a derivative with respect to $\Phi \equiv \Theta_d$, so that $\dot{\Theta}_d = 1$. The Euler-Lagrange equations are
\begin{equation}
	\frac{\partial \mathcal{L}}{\partial \Theta_i} - \frac{d}{d\Phi} \left( \frac{\partial \mathcal{L}}{\partial \dot{\Theta}_i} \right) = 0 \, ,
\end{equation}
where $\mathcal{L}$ is the integrand of (\ref{ddimlength}). We find that  
\begin{equation}\label{c4c4}
	\begin{split}
		\frac{\partial \mathcal{L}}{\partial \dot{\Theta}_i} &= \frac{\dot{\Theta}_i}{\mathcal{L}} \prod_{m=1}^{i-1} \cos^2 \Theta_{m} \,,  \\
		\frac{\partial \mathcal{L}}{\partial \Theta}_i &= - \frac{\sin \Theta_i \cos \Theta_i}{\mathcal{L}} \left(\prod_{n=1}^{i -1} \cos^2 \Theta_n  \right) \left( \sum_{a=i+1}^{d} \left( \prod_{m = i +1}^{a-1} \cos^2 \Theta_m \right) \dot{\Theta}_a^2 \right) \, .
	\end{split}
\end{equation}
with $ i = 1, \ldots, d -1$.  It can be shown that 
\begin{equation}
	\frac{d}{d \Phi} \left( \frac{\partial \mathcal{L}}{\partial \dot{\Theta}_i} \right) =   \left( \frac{\ddot{\Theta}_i}{\mathcal{L}} - \frac{\dot{\Theta}_i}{\mathcal{L}^2} \frac{d \mathcal{L}}{d\Phi} -  \frac{2 \dot{\Theta}_i}{\mathcal{L}} \sum_{j=1}^{i-1} \tan \Theta_j  \dot{\Theta}_j \right) \prod_{n=1}^{i-1} \cos^2 \Theta_n 
	\, . 
\end{equation}
The equation of motion will therefore have an overall factor of $\prod_{n=1}^{i -1} \cos^2 \Theta_n$. We will choose the endpoints of the geodesics to lie at $\Theta_i (\Phi_1)= 0 $ and $\Theta_i (\Phi_2)= 0 $ for all $ i = 1, \ldots , d-1$. Therefore, although it appears that $\Theta_i = \pm \tfrac{\pi}{2}$ would solve the equation of motion for $i = 2, \ldots d-1$, this solution does not obey the boundary conditions and so we neglect it. For $i=1$ this product evaluates to $1$, and so we can focus on the equation without this overall factor. Therefore, the equation of motion is 
\begin{multline} \label{ddimeom}
 \bigg(\sin \Theta_i \cos \Theta_i  \left( \sum_{a=i+1}^{d} \left( \prod_{m = i +1}^{a-1} \cos^2 \Theta_m \right) \dot{\Theta}_a^2 \right) 
	+  \ddot{\Theta}_i - \frac{\dot{\Theta}_i}{\mathcal{L}} \frac{d \mathcal{L}}{d\Phi} -  2 \dot{\Theta}_i \sum_{j=1}^{i-1}  \tan \Theta_j  \dot\Theta_j\bigg)  = 0 \, ,
\end{multline}
where 
\begin{equation}
	\frac{d \mathcal{L}}{d\Phi} = 
	\frac{\partial \mathcal{L}}{\partial \dot{\Theta}_i} \ddot{\Theta}_i+
	\frac{\partial \mathcal{L}}{\partial \Theta}_i
	\dot{\Theta}_i  \,,
\end{equation}
with the various terms above defined in equation \eqref{c4c4}. Therefore, we can see that $\Theta_i = 0$, for $i = 1 ,\ldots d-1$ is a solution to equation (\ref{ddimeom}). 

%% file: sections/appdetails.tex
\begin{figure}[H]
        \centering
        \includegraphics[height=5cm]{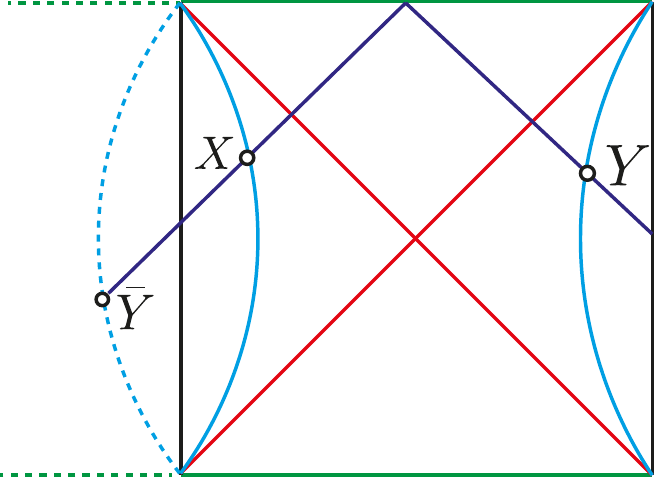}
\caption{Illustration of the null rays involved in fixing the critical static time $t_c$, after which symmetric geodesics between the two stretched horizons do not exist anymore.}\label{fig:transitiontc}
\end{figure}

It is useful to relate the global and static coordinates in embedding space. These are related to the three-dimensional Minkowski coordinates by
\begin{eqnarray}
		X_0 =& \sinh T\qquad &= \sqrt{1 - r^2} \sinh t \, , \nonumber\\   
		X_1 =& \cos \varphi \cosh T &= r\, ,   \\
		X_2 =& \sin \varphi \cosh T &= \sqrt{1-r^2} \cosh t \,  .\nonumber \label{Global_Static_rels}
\end{eqnarray}
Rearranging, we can write the global coordinates in terms of the static ones as
\begin{align}
T(r,t) &= \text{sinh}^{-1} \left(\sqrt{1-r^2} \sinh t  \right) \,, \\
\varphi(r,t)  &= \cos^{-1} \left( \frac{r}{\sqrt{\cosh^2 t - r^2\sinh^2 t }} \right) . \label{T_phiSP}
\end{align}
It is clear from this that the points opposite each other with a fixed $r = r_{st}$ will have the same $T_0$ and their angles will be at $\varphi_0$ and $- \varphi_0$. This gives the dS invariant distance written in the main text,
\begin{equation} 
    P_{X,Y}^{st} = -(X_0)^2+(X_1)^2-(X_2)^2 = \left(r_{st}^2-1\right) \cosh 2 t+r_{st}^2 \,.
\end{equation} 
We can also recover the critical time from geometric arguments. Starting from $t=0$ and increasing in time, note that there are two final geodesics at a time $|t|= t_c$ that are almost null. After this, there are no more real geodesics. Let us compute $P_{X,Y}$ for the endpoints of the final geodesics. Recall that in $d=2$ the full Penrose diagram is two copies of the usual square. If we call $Y$ the point on the right of the Penrose diagram, it is easy to see that its antipodal point $\bar{Y}$ will be outside the Penrose diagram and will be null separated to $X$, the left endpoint of the geodesic, see figure \ref{fig:transitiontc}. If $X$ is null separated from the antipodal point of $Y$, then the last geodesics have $P_{X,Y} = -1$. To compute $t_c$ we just need to use the metric in static coordinates (\ref{static_metric}), and find a null ray that passes through the point $(r=0, t=0)$. That ray intersects $r=r_{st}$ at exactly the $t_c$ given by (\ref{critical_t}). So, we have recovered this time scale from a geometric point of view.

